# Energy transfer between localized emitters in photonic cavities from first principles

Swarnabha Chattaraj [1†] and Giulia Galli[1,2,3*]

[1] Materials Science Division, Argonne National Laboratory, Lemont, Illinois 60439, USA
[2] Pritzker School of Molecular Engineering, University of Chicago, Chicago, Illinois 60637, USA
[3] Department of Chemistry, University of Chicago, Chicago, Illinois, 60637, USA
[†]schattaraj@anl.gov, [*]gagalli@uchicago.edu

**ABSTRACT**. Radiative and nonradiative resonant couplings between defects are ubiquitous phenomena in photonic devices used in classical and quantum information technology applications. In this work we present a first principles approach to enable quantitative predictions of the energy transfer between defects in photonic cavities, beyond the dipole-dipole approximation and including the many-body nature of the electronic states. As an example, we discuss the energy transfer from a dipole like emitter to an F center in MgO in a spherical cavity. We show that the cavity can be used to controllably enhance or suppress specific spin flip and spin conserving transitions. Specifically, we predict that a ~10 to 100 enhancement in the resonant energy transfer rate can be gained in the case of the F center in MgO at ~10nm distances from a dipolar source, using rather moderate cavity with quality factor Q~400. We also show that a similar suppression in the transfer rate can be achieved by off-tuning the cavity resonance relative to the emitter transition energy. The framework presented here is general and readily applicable to a wide range of devices where localized emitters are embedded in micro-spheres, core-shell nanoparticles, and dielectric Mie resonators. Hence, our approach paves the way to predict how to control energy transfer in quantum memories and in ultra-high density optical memories, and in a variety of quantum information platforms.

## I. INTRODUCTION.

Coupling and energy transfer processes between localized quantum emitters in solids, in particular near field nonradiative resonant energy transfer (NRET) [1-6], are relevant phenomena for various technological applications, e.g. classical photonic devices and quantum memories and networks. Controlling energy transfer processes can provide a way to entangle distinct quantum emitters suitably integrated in solid state nanophotonic devices and further enable entangling “swap” operations for quantum memories and networks [7-10]. In addition, NRET processes are known to lead to spectral diffusion and dephasing of optical and spin transitions resulting in decoherence pathways [8,11,12]. Thus, a quantitative understanding of NRET between localized defects in solid state photonics platform is of importance for the realization of classical and quantum optical devices.

The description of radiative and nonradiative energy transfer processes at arbitrary length-scale, has been unified [4,5, 13-15] within the framework of quantum electrodynamics (QED) where energy transfer is treated as transfer of virtual photons. In the dipolar limit, the QED framework has been extended to include the effect of inhomogeneous media on energy transfer and describe, for example, plasmonic resonances [16,17]. However, the design of devices involving realistic materials. requires (a) going beyond the dipole approximation to account for distributed electron states or current densities [18], (b) including many-body state transitions, and (c) accounting for spin and orbital states interacting with the electric and magnetic fields of the photons.

To this end, we have recently proposed a general framework [6] that couples first-principles electronic structure theories with quantum electrodynamics to incorporate the many body nature and spin degrees of freedom of the states of the localized emitters in the description of radiative and nonradiative energy transfer processes, beyond the two-level system assumption and the dipole approximation. Here, we extend our framework to include inhomogeneity in the dielectric medium in which the emitters are embedded and provide a theoretical platform to understand energy transfer processes from first principles and investigate many photonic device platforms of interest.

The building unit of photonic devices is an ensemble of localized quantum emitters controllably coupled amongst themselves, and with photonic nanostructures for efficient photon extraction, propagation, and interference [19,20]. These nanostructures include cavities [19-22], waveguides [19], and nano-antennas [23-25], with photonic cavities and nano-antennas being used, e.g. to enhance photon emission rate and emission directionality from embedded solid-state emitters. The presence of a photonic cavity also offers a way to dynamical tuning. For example, in photonic devices the strength of light-matter coupling between embedded quantum emitters and photonic modes can be controlled by thermal, electro-optic, or piezoelectric tuning [26-28] of cavity resonances. Designing modes of photonic nanostructures can also facilitate entanglement between quantum emitters [29-33] with the purpose of eventually building quantum networks and quantum simulation platforms [29,30]. Further, photonic structures are important in the context of ultra-high density optical memories [6] realized using a large ensemble of narrow-band deep level emitters. Recently we proposed that ultra-high density atomic memories [6] can be realized by optically addressing each individual emitter in an inhomogeneous ensemble so as to enable the transfer of excitations to a nearby trap defect, as shown in Fig. 1a. For illustrative purposes, the photonic cavity in Fig. 1(a) is a vertical Fabry-Perot cavity, but other implementations [13], including photonic crystals, micropillars, Mie resonances, and micro-ring whispering gallery resonators are also viable. In such systems, tuning a narrow band cavity to a specific set of quantum emitters can be used to activate the energy transfer process from the emitters to nearby defect states and therefore the cavity mode can be used to control the memory write process, as illustrated in Figure. 1(b). Thus, developing a quantitative predictive model to address near field energy transfer in inhomogeneous dielectric media—specifically nanophotonic cavity structures, is critically important.

Historically, interacting emitters mediated by a photonic cavity has been described by a Travis-Cumming model [34], where the cavity is assumed to be a narrow band comprising of a set of discrete delta-function photon modes. This assumption is adopted in most polaritonic chemistry [35-39] studies, where one is interested in polaritonic modes of electronic systems embedded in strongly coupled plasmonic, or ultra-high Q Fabry-Perot cavity with very high relative coupling strength (coupling energy/transition energy) up to ~ 0.1. In this regime, full self-consistency of the photonic mode and electronic structure is required [40], and the system size is therefore limited to nanometer scale. Our regime of interest is instead that of micron-scale photonic devices relevant to quantum information applications, networks, and communication platforms, where the photonic cavities are weakly coupled and possess at most a relative coupling strength of ~$10^{-4}$; in this case the discrete photonic mode approximation is not accurate. More importantly, discrete photon modes are insufficient to describe near field energy transfer phenomena, where the energy transfer is mediated by a broad spectrum of virtual photon modes [6, 41] whose energy broadening is controlled by the uncertainty timescale of photons of short-lived nature.

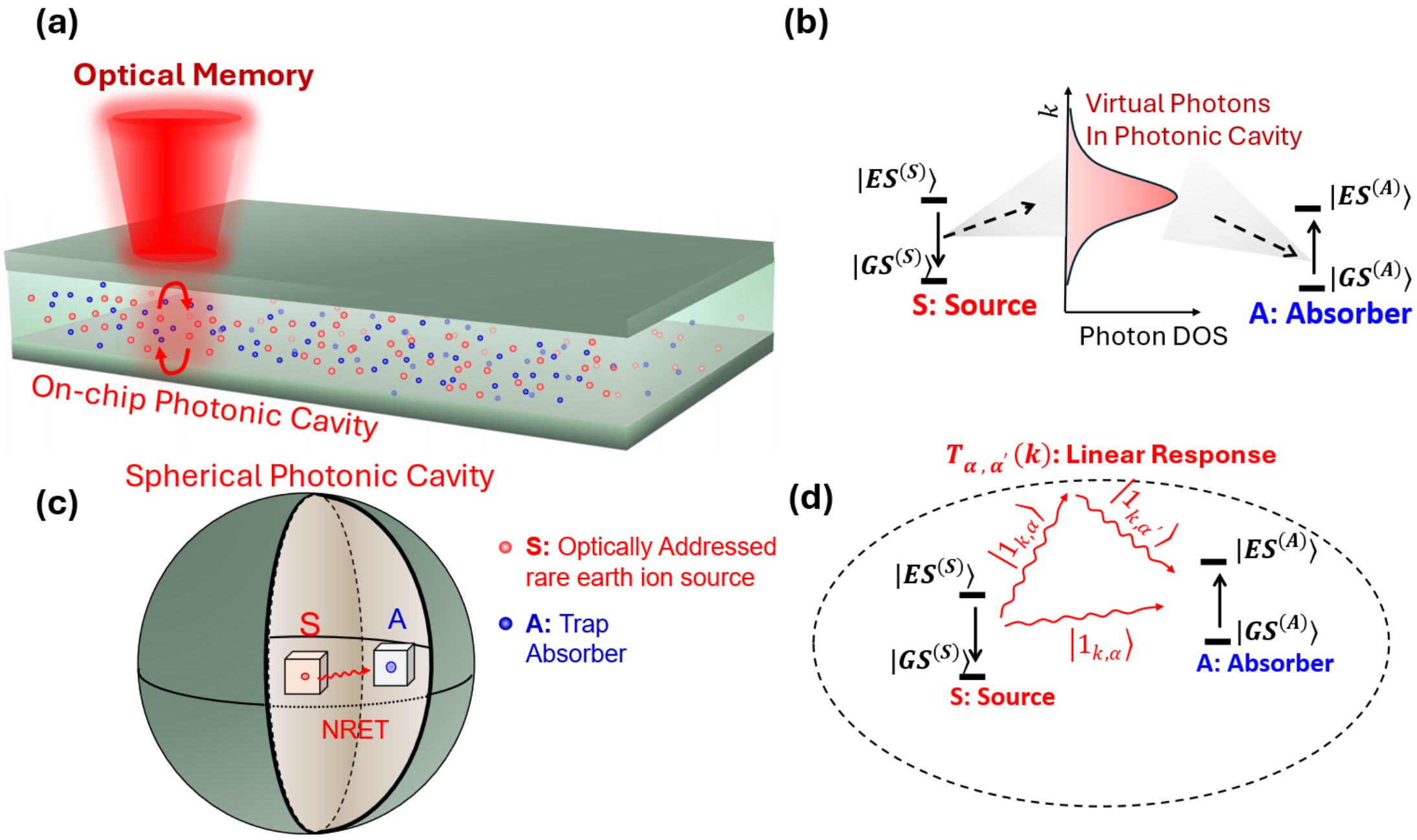

FIG 1. (a) Ensemble of source (e.g. rare earth ions) and absorbers (defect traps) in a solid-state photonic cavity used to build ultra-high density optical memories [6]. The NRET (nonradiative resonance energy transfer) from the source to the trap, mediated by the cavity mode provides the basic functional write-process. (b) Cavity mediated energy transfer. The virtual photon participating in the NRET process is localized in the cavity mode. Here $|GS^{(S/A)}\rangle$ and $|ES^{(S/A)}\rangle$ represent the ground state and the excited state of the source (S) and the absorber (A) respectively. (c) Example of a spherical cavity representing, e.g. a nanoparticle or a microparticle (d) Schematic representation of the effect of the cavity, treated with linear response, on energy transfer. The virtual photon is expanded in the basis denoted as $|1_{k,\alpha}\rangle$ where $k$ represents the wavenumber and $\alpha$ contains all other quantum numbers. The scattering matrix (see equation (14)) is denoted as $T_{\alpha,\alpha'}(k)$ .

Approaches that account for the broad continuum of photon modes in photonic nanostructures are predominantly based on the dipole-dipole approximation, with the emitters approximated as ideal two-level systems [31-33, 42-46]. Within the electric dipole approximation, the coupling matrix element between emitters can be expressed as $M = \bar{p}_S \cdot \bar{\bar{G}}(\omega;\ \bar{r}_S, \bar{r}_A) \cdot \bar{p}_A$ [31], where $\bar{p}_{S/A}$ are the point electric dipole approximating the transitions at the source (S)/absorber (A), and $\bar{\bar{G}}$ represents the classical electromagnetic Green's tensor defined by the Maxwell wave equation $\bar{\nabla} \times \bar{\nabla} \times \bar{\bar{G}}(\bar{r}, \bar{r}', \omega) - k^2 \bar{\bar{G}}(\bar{r}, \bar{r}', \omega) = \frac{k_0^2}{\epsilon_0} \bar{\bar{I}} \delta(\bar{r} - \bar{r}')$; here $k$ and $k_0$ represent the wavenumber of the photon in the dielectric medium and in vacuum respectively, and $\epsilon_0$ is the vacuum permittivity. The change of the dipole oscillator strength for the individual emitters is then given by the diagonal component of $\bar{\bar{G}}(\omega;\ \bar{r}_S, \bar{r}_A)$- i.e. the photon local density of states (LDOS) projected along a specific direction $\hat{p}_S$ of the source dipole and is given by $\rho_{LDOS}(\omega) = \frac{6\epsilon_0}{\pi\omega} Im\left(\hat{p}_S \cdot \bar{\bar{G}}(\bar{r}_S, \bar{r}_S, \omega) \cdot \hat{p}_S\right)$[24]. This formulation however only applies to point-like dipole sources [31] and multipolar modes, when considering the non-local components of the Green function [47], but it cannot be readily generalized to transitions between many electron states that have in general multireference character. Hence, to allow for the design of devices involving realistic emitters, with many-body electronic states, the dipole approximation can no longer be applied, and a more general theoretical framework is necessary.

Here we present a framework where we treat the source and the absorber with first principles many-body theories; a continuum of the virtual photon modes is accounted for by a quantized multipolar basis, and the presence of a cavity is described using linear response theory. We use the minimal coupling framework [6, 44, 48] and a Pauli Hamiltonian to account for the interaction of light with the many-body electronic states, and we include orbital and spin degrees of freedom. The framework presented here is generalized to various geometries by choosing a suitable basis for the virtual photon modes, and applicable to varied combinations of localized emitters representing a source and an absorber. As a specific example, and to compare with our previous work [6], we show the effect of a spherical cavity on the energy transfer process between an ideal dipole source and a F center in MgO, acting as an absorber.
.

The rest of the paper is organized in the following way. In Section II, we present the theoretical framework to account for the effect of arbitrary photonic structures on the radiative and nonradiative energy transfer process between a generalized source and an absorber at arbitrary length scales. In Section III, we present the example of the optical absorption of the F center in MgO from electric or magnetic dipole-like sources, embedded in a spherical nanoparticle forming a cavity. Finally in Section IV we present our conclusions.

## II. METHODS

We build upon our previous work [6] where we presented a generalized framework to describe the energy transfer in homogeneous dielectric media within a weak coupling regime, using a quantized multipolar basis representing the virtual photon modes participating in the energy transfer process. In the presence of the cavity, the Hamiltonian is:

$$H = H_S + H_A + H_{Field} + H_{int} + H_{1Cav} \quad (1)$$

The expressions of $H_S, H_A, H_{Field}$ are the same as in Ref. [6] and they are summarized below. The Hamiltonians $H_S$ and $H_A$ represent effective Hamiltonians of the isolated source and the absorber. Each of them can be defined over a Hilbert space that is spanned by Slater determinants of $N_{S/A}$ electrons occupying a chosen set of active orbitals $\left\{\left|\phi_i^{(S/A)}\right\rangle\right\}$. Both Hamiltonians $H_S$ and $H_A$ may be approximated by Kohn-Sham Hamiltonians [49-51], or with effective Hamiltonians using quantum embedding theories and many body perturbation theories [52-56]. To describe the NRET process, we consider the ground (GS) and a given excited state (ES) of the source and absorber, which in general can have a multireference nature and in second quantization can be represented as $\left|GS^{(S/A)}\right\rangle = \sum_{\substack{i\epsilon\, occ \\ j\epsilon\, unocc}} \alpha_{ij,S/A}^{(GS)} c_j^{(S/A)\dagger} c_i^{(S/A)} \left|D^{(S/A)}\right\rangle$ and $\left|ES^{(S/A)}\right\rangle = \sum_{\substack{i\epsilon\, occ \\ j\epsilon\, unocc}} \alpha_{ij,S/A}^{(ES)}\ c_j^{(S/A)\dagger} c_i^{(S/A)} \left|D^{(S/A)}\right\rangle$ , respectively, where $|D\rangle$ is a Slater determinant built from the first

filled N orbitals, i.e., $\left|D^{(S/A)}\right\rangle = \prod_{i=1}^{N} c_i^{(S/A)\dagger}\,|0\rangle$. Here $c_i^{(S/A)}$ denotes the annihilation operator of an electron in the single particle orbital $\left|\phi_i^{(S/A)}\right\rangle$. The field Hamiltonian can be simply expressed as $H_{Field} = \sum_{k,\alpha} \hbar\omega_k \left(a_{k,\alpha}^{\dagger} a_{k,\alpha} + \frac{1}{2}\right)$, where $a_{k,\alpha}^{\dagger}$ is the creation operator of a photon in the mode $\{k, \alpha\}$; $k$ denotes the wavenumber of the photon and $\alpha$ denotes all other degrees of freedom specifying a mode. To represent the photon modes $\left|1_{k,\alpha}\right\rangle$ we use a complete eigen basis of the Maxwell equations in a homogeneous dielectric medium. Depending on the specific problem, this basis can be chosen as plane waves [5], spherical waves [6], or cylindrical waves [57]. For example, in an homogeneous bulk, the vector spherical harmonics are a convenient localized basis, as demonstrated in our earlier work [6], for which $\alpha = \{L, Jz, P\}$ where $L$ is the orbital angular momentum, $J_z$ the z-projected total angular momentum with integer values from -L to L, and P is the parity of the photon mode with values from {-1, 1}.

We use the same interaction Hamiltonian (nonrelativistic Pauli form) as in Ref. [6]:

$$H_{int} = \sum_{E=S,A} \sum_{i=1}^{N_E} \left[ -\frac{e\,\overline{\boldsymbol{p}}_i \cdot \overline{\boldsymbol{A}}}{2m_0} - \frac{e\,\overline{\boldsymbol{A}} \cdot \overline{\boldsymbol{p}}_i}{2m_0} + eA_0 + \mathrm{g}\frac{e\hbar}{2m_0} \overline{\boldsymbol{\sigma}}_i \cdot \overline{\boldsymbol{\nabla}} \times \overline{\boldsymbol{A}} \right] \tag{2}$$

Here $\overline{\boldsymbol{p}}_i$ represents the momentum operator of the $i$th electron, $m_0$ the rest mass of the electron, $g$ the gyromagnetic factor, and $\overline{\sigma}_i$ are Pauli matrices. We have neglected the elastic scattering term $\bar{A} \cdot \bar{A}$ term as it does not contribute to the second order perturbative NRET process, but it can be included if needed when incorporating higher order terms. We note here that- because of the equivalence between the momentum operator ($\overline{\boldsymbol{p}}$) and the current density operator ($\overline{\boldsymbol{j}}$) of the electronic states, the $\overline{\boldsymbol{A}} \cdot \overline{\boldsymbol{p}}$ term is equivalent to the the $\overline{\boldsymbol{A}} \cdot \overline{\boldsymbol{j}}$ term used in other works [18,58].

Similar to Ref.6, in this work we adhere to the minimal coupling picture [44] in the Coulomb gauge- thus enabling the straightforward use of the electronic orbitals obtained using first principles theories. The effect of the dielectric constant of the medium is contained in expressions of the field $\overline{\boldsymbol{A}}$ [Appendix A, Ref. 6] and thus does not explicitly appear in equation (2). We also note that the $\overline{\boldsymbol{A}} \cdot \overline{\boldsymbol{p}}$ term is equivalent to the more commonly used $\overline{\boldsymbol{E}} \cdot \overline{\boldsymbol{r}}$ term in resonant conditions [59], and in cases where the electron Hamiltonian involves only local interactions [6]. The latter assumptions is often not satisfied in density functional theory-based calculations. Hence, using the interaction Hamiltonian of equation (2) provides direct compatibility with first principles electronic structure calculations.

The difference between the formulation presented here and in Ref. [6] is the presence of a cavity, described by the Hamiltonian $H_{1Cav}$, which accounts for the inhomogeneity of the dielectric medium resulting in the scattering between the states of the eigen basis $\left|1_{k,\alpha}\right\rangle$ of the homogeneous medium:

$$H_{1Cav} = \sum_{k,\alpha,\alpha'} h_{\alpha,\alpha'}(k) \left|1_{k,\alpha'}\right\rangle\left\langle 1_{k,\alpha}\right| \tag{3}$$

The off-diagonal terms ($\alpha \neq \alpha'$) represent the scattering between different multipolar modes, whereas the diagonal terms ($\alpha = \alpha'$) represent an additional dispersion introduced by the photon scattering. Here we assume that the scattering is elastic, i.e. nonlinear media are not included in our description, and thus $k$ remains a good quantum number in equation (3).

In the absence of $H_{1Cav}$, within second order perturbation theory, the probability amplitude for the energy transfer process is [6]

$$c(t) = \frac{1}{\hbar^2} \int dk \sum_{\alpha} \frac{v_{k\alpha}^{(S)}\, v_{k\alpha}^{(A)}}{\omega_k - \omega_S} \left( \frac{e^{-i\omega_S t} - e^{-i\omega_A t}}{\omega_A - \omega_S} - \frac{e^{-i\omega_k t} - e^{-i\omega_A t}}{\omega_A - \omega_k} \right) \tag{4}$$

where $v_{k,\alpha}^{(A)} = \left\langle ES^{(A)} \left| \widetilde{H}_{int} \right| GS^{(A)}, 1_{k,\alpha} \right\rangle / \sqrt{\Delta k}$, $v_{k,\alpha}^{(S)} = \left\langle GS^{(S)}, 1_{k,\alpha} \left| \widetilde{H}_{int} \right| ES^{(S)} \right\rangle / \sqrt{\Delta k}$ are the matrix elements corresponding to photon absorption and emission processes in the interaction picture, $\omega_S$ and $\omega_A$ represent the transition energies at the source and the absorber respectively, and $\omega_k$ the energy of the photon. In the following we show how equation (4) is modified in the presence of a cavity.

### A. Linear response of the cavity

In the single photon limit, the scattering of the photon modes can be solved by employing Maxwell equations. To do so, we initialize a specific multipolar source of an electromagnetic wave at the location of the source, and we solve for the scattering into other multipolar modes, resulting in a scattering matrix $T_{\alpha,\alpha'}(k)$, as shown in Fig. 1(d). The matrix $T_{\alpha,\alpha'}(k)$, can be obtained either analytically for spherically symmetric cases (see Appendix A) or using numerical techniques, e.g. the finite difference time domain method, or the finite element method for more generic structures. The Green function of the photon can then be represented in the multipolar basis as

$$\hat{D}_{\alpha,\alpha'}(\omega) = \sum_{k,\alpha,\alpha'} \frac{2\omega_k}{\omega^2 - \omega_k^2}\left[\delta_{\alpha,\alpha'}\left|1_{k,\alpha'}\right\rangle\left\langle 1_{k,\alpha}\right| + T_{\alpha,\alpha'}(k)\left|1_{k,\alpha'[J]}\right\rangle\left\langle 1_{k,\alpha}\right|\right] \qquad (5)$$

Note that here the radial function of the multipoles is assumed to be a spherical Hankel function. The scattered wave however is slowly varying at the source and thus can be represented with a spherical Bessel function, indicated by the subscript [J].

The linear response of the cavity affects the energy transfer process in two ways. First, in the presence of the cavity, the transition corresponding to photon emission from the source is dressed with the photon mode; in the weak coupling regime this leads to the Purcell effect and to a small shift in the transition energy. Second, the propagation of the photon from the source to the absorber is affected. Both effects need to be accounted for to obtain a complete picture of the modification of the energy transfer process by the photonic cavity. They are discussed next.

### *1. Effect of the cavity mode on the source oscillator strength*

We start by considering the effect of the vacuum state (zero photon occupation) in the cavity on the transition at the source. One can think of the process as emission and immediate reabsorption of the photon, represented by the Dyson sequence shown in Fig. 2(a). Truncation of this sequence to only the first term defines the weak coupling limit [45,60] which is the relevant one for most nanophotonic device platforms [19], as mentioned in the Introduction. In this limit, self-energy can be evaluated as:

$$\Sigma \approx \frac{i\pi n_i}{\hbar c}\sum_{\alpha}\sum_{\alpha'}\left[\delta_{\alpha,\alpha'} v^{(S)*}_{k_S\alpha}\, v^{(S)}_{k_S\alpha'} + T_{\alpha,\alpha'}(k)\, v^{(S)*}_{k_S\alpha}\, v^{(S)}_{k_S\alpha'}\right] \qquad (6)$$

where $k_S$ corresponds to the wavevector of the source transition energy. The real part of $\Sigma$ represents the Lamb shift and the imaginary part represents the radiative decay rate [60]. From equation (6) we obtain the radiative decay time $T_1 = \frac{2\hbar}{Im(\Sigma)} = \frac{2c\hbar^2}{\pi n_i} Real\left(\sum_\alpha \sum_{\alpha'}\left[\delta_{\alpha,\alpha'} v^{(S)*}_{k_S\alpha}\, v^{(S)}_{k_S\alpha'} + T_{\alpha,\alpha'}(k) v^{(S)*}_{k_S\alpha} v^{(S)}_{k_S\alpha'}\right]\right)^{-1}$, and the Lamb shift $Re(\Sigma) = Imag\left(\frac{\pi n_i}{\hbar c}\sum_\alpha \sum_{\alpha'}\left[\delta_{\alpha,\alpha'} v^{(S)*}_{k_S\alpha}\, v^{(S)}_{k_S\alpha'} + T_{\alpha,\alpha'}(k) v^{(S)*}_{k_S\alpha} v^{(S)}_{k_S\alpha'}\right]\right)$.

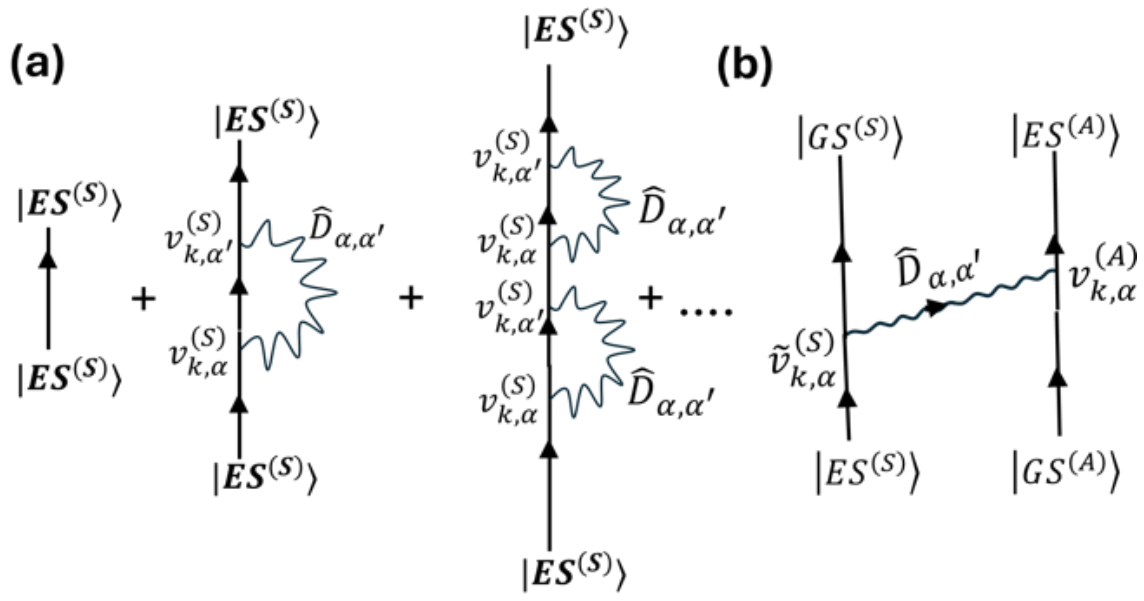

FIG 2. (a) Diagrammatic expansion of the source transition being dressed by the vacuum cavity mode (see text). Truncating at the first order term defines the weak coupling limit [45,60]. (b) Diagram showing the energy transfer process with the photon propagator modified by the cavity. Here $|GS^{(S/A)}\rangle$ and $|ES^{(S/A)}\rangle$ are the ground state and the excited state of the source (S) and the absorber (A) respectively and $v^{(S/A)}_{k,\alpha}$ denote the matrix elements corresponding to photon emission and absorption processes (see text). $\hat{D}_{\alpha,\alpha'}$ denotes the photon propagator in the presence of the photonic cavity as shown in equation (5).

In the absence of the cavity, we have

$$\Sigma_0 = \frac{i\pi n_i}{\hbar c}\sum_{\alpha}\left|v^{(S)}_{k_S\alpha}\right|^2 \qquad (7)$$

which results in the radiative decay rate time of the source in an uniform dielectric medium $T_1^{(0)} = \frac{2\hbar}{Im(\Sigma_0)} = \frac{2c\hbar^2}{\pi n_i \sum_\alpha \left|v^{(S)}_{k_S\alpha}\right|^2}$. The Purcell enhancement is defined as the enhancement of the radiative decay rate of the source transition in the presence of the cavity, compared to the case where the source is in an infinite bulk material with no cavity. Thus, from equation (6) and (7) we can obtain a general expression of the emission rate enhancement:

$$F_p = \frac{Im(\Sigma)}{Im(\Sigma_0)} \qquad (8)$$

We note that equation (8) is a generalized version of the Purcell enhancement of an electric dipole emitter that can be expressed using the classical electromagnetic Green's function as $F_p \propto Im\left(G(\omega_s; \bar{r}_S, \bar{r}_S)\right)$. Equation (8) is general and holds for realistic transitions involving many-body electronic states where the transitions cannot be approximated as an electric dipole.

The formulation in equation (8) captures the effect of arbitrary transitions between multireference electronic states, as modified by the electric and magnetic resonances of the medium that can be present simultaneously. The effect of the electric dipolar, magnetic dipolar, or any other type of cavity multipolar interactions are automatically included in the sum over all $\alpha's$. Including all interactions is important; for example, in dielectric resonators, such as Mie resonators with both electric and magnetic resonances [25], source and absorber transitions can be both of an electric and magnetic nature in the same spectral range. Thus, our framework provides for the first time a way to estimate the effect of generic nanophotonic environments on the electronic transitions of realistic spin defects, paving the way to the design of optical and quantum memory devices.

Next, we use how the photon mediated energy transfer process is modified in the presence of the cavity.

### *2. Energy transfer mediated by the cavity*

To include the effect of the linear response of the cavity onto the transfer of the photon from the source to the absorber (see Fig. 2(b)), we write the energy transfer probability amplitude as:

$$c(t) = \frac{1}{\hbar^2}\int dk \sum_{\alpha}\sum_{\alpha'}\Bigg[\frac{1}{\omega_k-\omega_S} \left(\delta_{\alpha,\alpha'} v_{k\alpha}^{(S)}\, v_{k\alpha'}^{(A)} + T_{\alpha,\alpha'}(k) v_{k\alpha}^{(S)}\, v_{k\alpha'[J]}^{(A)}\right)\left(\frac{e^{-i\omega_S t}-e^{-i\omega_A t}}{\omega_A-\omega_S} - \frac{e^{-i\omega_k t}-e^{-i\omega_A t}}{\omega_A-\omega_k}\right)\Bigg] \tag{9}$$

where we have used the photon propagator from equation 5. The enhancement of the photon LDOS at the location of the source does need to be accounted for separately since it is contained in the modified photon Green function [61]. The functions $v_{k\alpha'}^{(A)}$ and $\tilde{v}_{k\alpha}^{(S)}$ are smooth functions of $k$. Also, in the weak coupling regime where the cavity linewidth is much larger than $\frac{\hbar}{t}$, $T_{\alpha,\alpha'}(k)$ can be considered constant over the uncertainty energy $\frac{\hbar}{t}$ around the transition energy $k_S$. Thus the factor $\left(\delta_{\alpha,\alpha'} v_{k\alpha}^{(S)}\, v_{k\alpha'}^{(A)} + T_{\alpha,\alpha'}(k) v_{k\alpha}^{(S)}\, v_{k\alpha'[J]}^{(A)}\right)$ can be taken out of the k-integration resulting in a simpler approximate form:

$$c(t) = \frac{1}{\hbar^2}\sum_{\alpha}\sum_{\alpha'}\left(\delta_{\alpha,\alpha'} v_{k_S\alpha}^{(S)}\, v_{k_S\alpha'}^{(A)} + T_{\alpha,\alpha'}(k_S) v_{k_S\alpha}^{(S)}\, v_{k_S\alpha'[J]}^{(A)}\right)\int \frac{1}{\omega_k-\omega_S}\left(\frac{e^{-i\omega_S t}-e^{-i\omega_A t}}{\omega_A-\omega_S} - \frac{e^{-i\omega_k t}-e^{-i\omega_A t}}{\omega_A-\omega_k}\right)dk \tag{10}$$

Further, for large t, the last term can be neglected resulting in a simple contour integral [6] to give:

$$c(t) = \frac{2i\pi n_i}{\hbar c}\sum_{\alpha}\sum_{\alpha'}\left[\delta_{\alpha,\alpha'} v_{k_S\alpha}^{(S)}\, v_{k_S\alpha'}^{(A)} + T_{\alpha,\alpha'}(k_S) v_{k_S\alpha}^{(S)}\, v_{k_S\alpha'[J]}^{(A)}\right]\frac{\sin\left(\Delta\omega\,\frac{t}{2}\right)}{\Delta\omega} = \frac{2M\sin\left(\frac{\Delta\omega}{2}\,t\right)}{\hbar\Delta\omega} \tag{11}$$

where $\Delta\omega = \omega_A - \omega_S$, and the matrix element of the energy transfer process, $M$ is given by:

$$M = \frac{i\pi n_i}{\hbar c}\sum_{\alpha}\sum_{\alpha'}\left[\delta_{\alpha,\alpha'} v_{k_S\alpha}^{(S)}\, v_{k_S\alpha'}^{(A)} + T_{\alpha,\alpha'}(k_S) v_{k_S\alpha}^{(S)}\, v_{k_S\alpha'[J]}^{(A)}\right] \tag{12}$$

Equation (11) is to be compared to equation (4), valid in the homogeneous case. A key assumption here is that we only account for energy transfer to the second perturbative order, as indicated by the diagram shown in Fig. 2(b). This is a realistic assumption in many systems [62-63] and particularly relevant to optical memories (Fig. 1(a)) where the objective is to trap the excitation into a long-lived state at the absorber. Physically the trapping is enabled by the Stoke shift of the absorber's vertical transition due to the excited state relaxation process that prevents re-emission from the absorber.

## B. Spherically symmetric cavity

The formulation provided above is agnostic to the choice of the basis representing the virtual photon modes. As a specific example, here we consider a simple case where the source is located at the center of a spherical cavity (see Fig. 3a). The virtual photon modes are represented as multipolar modes centered at the center of the sphere. The spherical cavity can be realized using core-shell structures of micro- and nanoparticles, as discussed in appendix A. The shell of the nanoparticle provides an effective reflectivity, denoted as Γ. The photon modes inside the sphere are expressed as spherical waves - $\left|1_{\,k,\alpha[Z]}\right\rangle$. The sub-script Z denotes the type of Bessel function used to describe the photon and it can be J to represent standing waves, H1 (Hankel type 1) to represent radially outward propagating waves and H2 (Hankel of type 2) to represent radially inward propagating waves. A radially outward propagating photon (spherical Bessel H1) is reflected by the boundary of the cavity into a standing wave mode (spherical Bessel J). We employ Maxwell equations to solve for the electromagnetic fields with specific boundary conditions of continuity of tangential E and H fields at the dielectric interfaces (see Appendix A). If the spherical particle shares the same center with the

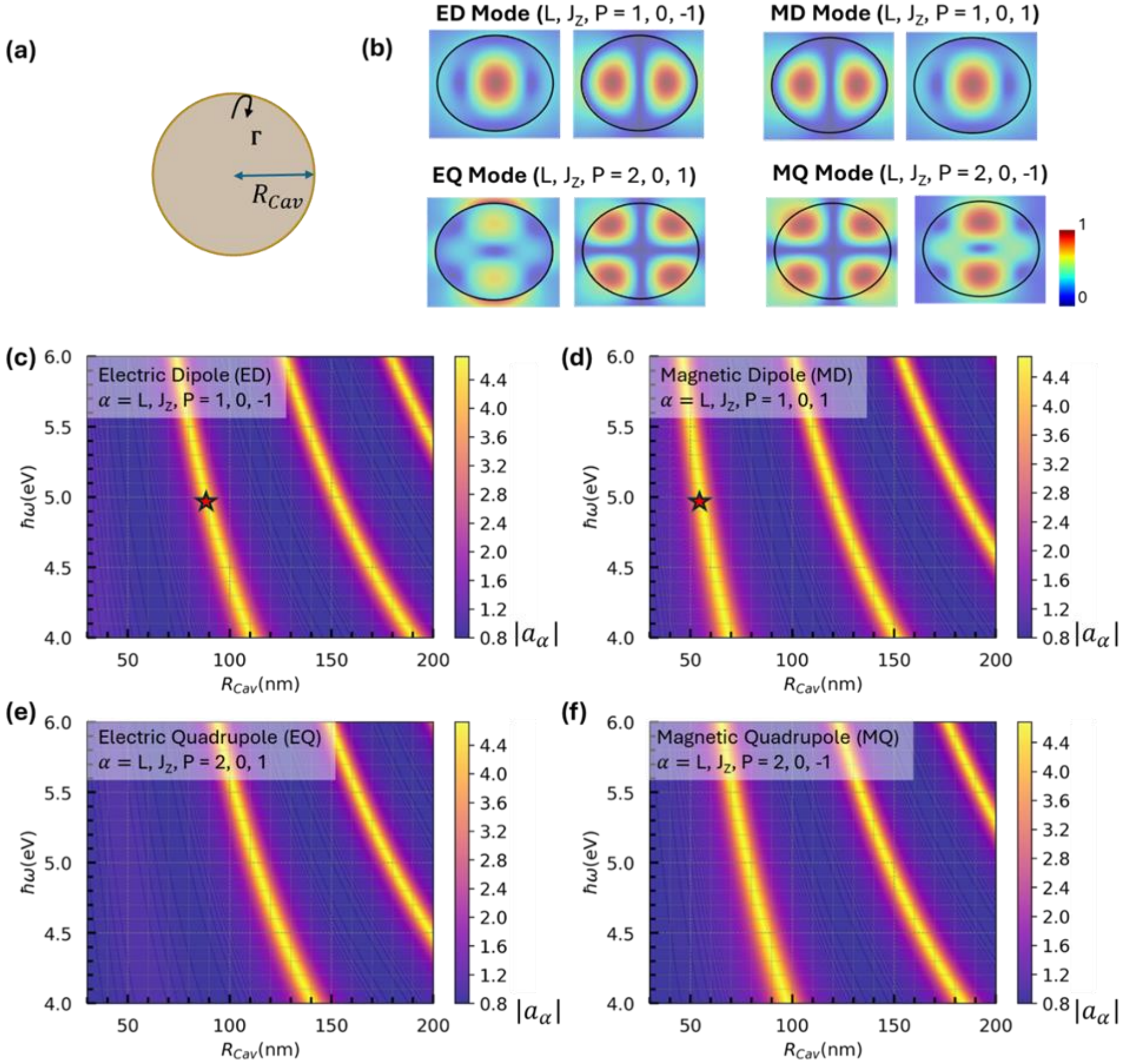

FIG 3. Response of a spherical cavity as a function of its size. The structure of the cavity is defined in (a). Panel (b) shows the distribution of the vector potential ($\bar{A}$, left panels) and magnetic field ($\bar{\nabla} \times \bar{A}$, right panels) inside the sphere for different modes including electric dipole, magnetic dipole, electric quadrupole, and magnetic quadrupole. Panels (c) to (f) show the response $a_\alpha$ as a function of photon energy and cavity radius for electric dipole (ED), magnetic dipole (MD), electric quadrupole (EQ), and magnetic quadrupole (MQ) modes corresponding to $\alpha = \{L = 1, J_Z = 0, P = -1\}$,$\{L = 1, J_Z = 0, P = 1\}$, $\{L = 2, J_Z = 0, P = 1\}$, and $\{L = 2, J_Z = 0, P = -1\}$, respectively. The different streaks correspond to modes of different radial orders. We investigated two values of $R_{Cav}$ as indicated by the red stars in panel (c) and (d), with $R_{Cav} = 55.2\ nm$ having a magnetic dipole resonance at 5 eV, and $R_{Cav} = 88.4\ nm$ having an electric dipolar resonance at 5eV (see text).

multipolar basis, the mode $\alpha$'s $(L, J_Z, P)$ are unaffected by the scattering process resulting in

$$T_{\alpha,\alpha'}(k) = a_\alpha(k)\,\delta_{\alpha,\alpha'} \qquad (13)$$

Under the assumption of a spherical cavity, the Purcell enhancement expression from equation (8) is simplified as:

$$F_p = \frac{\sum_\alpha \left|v^{(S)}_{k_S\alpha'[J]}\right|^2 \left(1 + Re\left(a_\alpha(k_S)\right)\right)}{\sum_\alpha \left|v^{(S)}_{k_S\alpha'[J]}\right|^2} \qquad (14)$$

The energy transfer matrix element in the spherical cavity approximation can be written as:

$$M_{NRET} = \frac{i\pi n_i}{\hbar c} \sum_\alpha v^{(S)}_{k_S\alpha}\left[v^{(A)}_{k_S\alpha} + a_\alpha(k_S)\, v^{(A)}_{k_S\alpha[J]}\right] \qquad (15)$$

Equations (12) and (15) provide a general framework to describe energy transfer processes for any arbitrary dielectric environment, where the response $T_{\alpha,\alpha'}(k)$ can be computed from the Maxwell equations. Thus, the above framework can be used to obtain the coupling and energy transfer between arbitrary emitters embedded in a nanophotonic device. Note that the effects of the modifications of the emission rate of the source by the cavity, and the source to absorber energy transfer have been expressed using matrix elements ($v^{(S)}_{ks\alpha}$ and $v^{(A)}_{ks\alpha}$) of the Pauli Hamiltonian with many-electron states, including orbital and spin degrees of freedom. Importantly, here we account for dipole-allowed, and dipole-forbidden transitions on the same footing, thus enabling the determination of the effect of a photonic nanostructure on electric dipole-forbidden transitions for arbitrary quantum emitters. The localized emitter can be, for example, native and implanted atomic defects including rare earth, deep level color centers, and distributed defects such as colloidal or epitaxial quantum dots.

## III. RESULTS

In this section, we present results for an exemplary system, where the absorber is an F center in MgO, and the source is an ideal electric or magnetic dipole.

The oxygen vacancy center in MgO has been extensively investigated both experimentally and theoretically over several decades, utilizing techniques such as optical absorption, photoluminescence, and electron spin resonance [64-67]. Experimental studies have identified the optical absorption of neutral F centers around ~5 eV, with emissions occurring at approximately ~2.3 eV and ~3 eV. The optical absorption is a result of a transition between a localized mid-gap s-type orbital ($|s\rangle$) and localized p-type orbitals ($|p_x\rangle, |p_y\rangle, |p_z\rangle$) just above the conduction band edge (see [6] and Appendix C for details). In the ground state configuration of the neutral F center, both spin states of the s-orbitals are filled resulting in a singlet ground state $|\,{}^1A_{1u}\,\rangle = |s_\uparrow, s_\downarrow\rangle$. The first excited singlet state can be written as $|\,{}^1T_{1u}\,\rangle = \frac{1}{\sqrt{2}}(|p_\uparrow, s_\downarrow\rangle + |s_\uparrow, p_\downarrow\rangle)$ whereas the three triplet excited states are: $\left|\,{}^3T_{1u,m_s=0}\right\rangle = \frac{1}{\sqrt{2}}(|p_\uparrow, s_\downarrow\rangle - |s_\uparrow, p_\downarrow\rangle)$, $\left|\,{}^3T_{1u,\ m_s=1}\right\rangle = |s_\uparrow, p_\uparrow\rangle$, and $\left|\,{}^3T_{1u,\ m_s=-1}\right\rangle = |\,p_\downarrow, s_\downarrow\rangle$. We calculate the electronic orbitals using Kohn Sham DFT (details in Appendix C, and Ref. 6). Note that a more accurate many-body approach [52-56, 67] can be readily implemented to better approximate the energy levels and the Slater determinants corresponding to the many-electron localized eigenstates. We then calculate the matrix elements between the singlet ground state $|\,{}^1A_{1u}\,\rangle$ and the singlet excited state $|\,{}^1T_{1u}\,\rangle$ or the triplet excited states $\left|\,{}^3T_{1u,m_s=0}\right\rangle$, $\left|\,{}^3T_{1u,m_s=1}\right\rangle$, *and* $\left|\,{}^3T_{1u,m_s=-1}\right\rangle$ in the following way:

$$\left\langle\, {}^1T_{1u}\left|H_{int}\right|GS\,, 1_{k,\alpha}\right\rangle = \frac{1}{\sqrt{2}}\langle p_\uparrow\left|H_{int}\right|s_\uparrow, 1_{k,\alpha}\rangle + \frac{1}{\sqrt{2}}\langle p_\downarrow\left|H_{int}\right|s_\downarrow, 1_{k,\alpha}\rangle \tag{16}$$

$$\left\langle\, {}^3T_{1u,m_s=1}\left|H_{int}\right|GS\,, 1_{k,\alpha}\right\rangle = \langle p_\uparrow\left|H_{int}\right|s_\downarrow, 1_{k,\alpha}\rangle \tag{17}$$

$$\left\langle\, {}^3T_{1u,\ m_s=-1}\left|H_{int}\right|GS\,, 1_{k,\alpha}\right\rangle = \langle p_\downarrow\left|H_{int}\right|s_\uparrow, 1_{k,\alpha}\rangle \tag{18}$$

$$\left\langle\, {}^3T_{1u,\ m_s=0}\left|H_{int}\right|GS\,, 1_{k,\alpha}\right\rangle = \frac{1}{\sqrt{2}}\langle p_\uparrow\left|H_{int}\right|s_\uparrow, 1_{k,\alpha}\rangle - \frac{1}{\sqrt{2}}\langle p_\downarrow\left|H_{int}\right|s_\downarrow, 1_{k,\alpha}\rangle \tag{19}$$

The radial function in the multipole photon mode used to obtain $v^{(A)}_{ks\,\alpha}$ or $v^{(A)}_{ks\,\alpha\,[J]}$ is chosen to be either spherical Hankel of type 1, or spherical Bessel J.

We demonstrate the effect of the cavity mode on the dipole allowed and dipole forbidden NRET processes using a spherical dielectric enclosed by a reflective surface. Quantum emitters embedded in nanoparticles have long been of interest due to the possibility of integrating them into a vast range of nanophotonic devices [68] and due to the interest in using a bottom-up synthesis approach. For sizes much smaller than the wavelength of the photon in the medium ($\sim\lambda/n_i$), the optical response is dominated by Rayleigh scattering providing an electric dipolar response. On the other hand, if the size of the nanoparticle is comparable to the wavelength, Mie resonances [25, 69-73] are allowed and they can provide both magnetic and electric responses of various orders, including dipolar, quadrupolar, and octupolar modes. For example, the magnetic dipole response is caused by the resonance of the electric displacement current in the dielectric medium within an enhanced magnetic field at the center of the nanoparticle. Such resonance is expected to favor orbital and spin forbidden transitions, given their magnetic dipolar nature. Thus, the interplay between the size of the nanoparticle governing the Mie resonance mode, and the nature of the many-body transition at the source and the absorber provide a rich space to investigate design principles for enhancing specific transitions partaking in the transfer process.

Mie resonances have been observed in dielectric building blocks of size $\sim\lambda/n_i$ , for various sizes and shapes including spheroidal, cylindrical, or rectangular, and for various material systems including oxides and semiconductors [69-73]. The quality factor of these resonances is typically limited to a few hundreds. However, a nanoparticle can be coated with a high index dielectric, or high conductivity metal layer in a core-shell structure to enhance the reflection from the surface, resulting in higher Q modes. Here we choose a spherical nanoparticle of radius $R_{Cav}$, and we assume a semi reflective coating at the surface with a reflectivity $\Gamma$ taken as a parameter, which is determined by the core-shell material combination, as shown in Appendix A.

For the spherical cavity shown in Fig. 3(a) we use the boundary conditions corresponding to a dielectric interface and we solve for the cavity response $a_{\alpha}(k)$ analytically (see Appendix A). Figure 3(b) shows the distribution of the vector potential and the magnetic field inside the nanoparticle for different values of $\alpha$- specifically electric dipole (ED) ($\alpha = \{L = 1, J_Z = 0, P = -1\}$), magnetic dipole (MD) ($\alpha = \{L = 1, J_Z = 0, P = 1\}$), electric quadrupole (EQ) ($\alpha = \{L = 2, J_Z = 0, P = 1\}$), and magnetic quadrupole (MQ) ($\alpha = \{L = 2, J_Z = 0, P = -1\}$) modes. Of particular interest are the ED and the MD modes since at the center of the nanoparticle, where the source is located, they exhibit an antinode of the vector potential $|\bar{A}|$ and of the magnetic field $|\bar{\nabla} \times \bar{A}|$, respectively.

By controlling the size of the nanoparticle, it is possible to control which specific mode is enhanced by the cavity at a desired transition energy. To illustrate the available design space, in Figure 3(c) to (f) we show the response $a_{\alpha}(k)$ corresponding to the ED, MD, EQ and MQ modes as a function of the cavity radius. We see that at any specific energy, with increasing nanoparticle radius, there are multiple peaks in the response (indicated by the multiple yellow streaks). Figure 3(c) and 3(d) indicate the specific size of the nanoparticle one should choose to enhance electric dipole and magnetic dipole transitions. Based on these results, we chose two specific sizes – $R_{Cav}$= 55.2 nm and 88.4 nm - as indicated by the red stars - that are analyzed in detail below. At these two sizes, we have a magnetic dipole and an electric dipole resonance, respectively, at 5 eV, the source transition energy- and in the following we investigate the nonradiative resonant energy transfer processes from an electric dipole source / magnetic dipole source to the oxygen vacancy in MgO as an absorber.

### A. Purcell enhancement spectra

In Figure 4, we show how the cavity mode modifies the oscillator strength of the source- which can be either an electric dipole ($\alpha = \{L = 1, J_Z = 0, P = -1\}$) or a magnetic dipole ($\alpha = \{L = 1, J_Z = 0, P = 1\}$). The generalized Purcell enhancement is computed using equation (8) and (6) and using the spherical

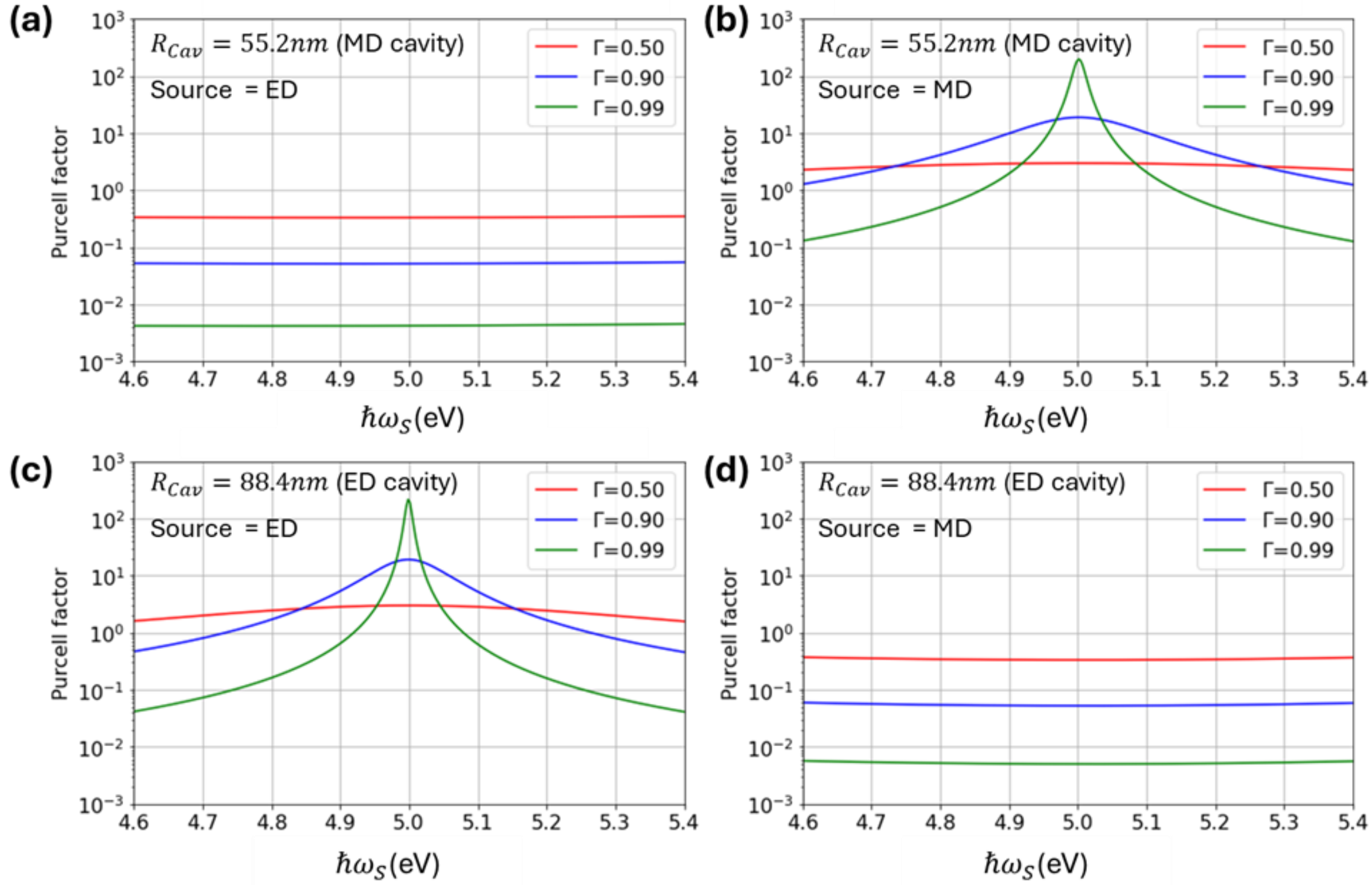

FIG 4. Generalized Purcell enhancement spectra (equation (14)) for electric (ED) / magnetic dipole (MD) source embedded in cavity with electric or magnetic dipolar modes. (a) ED source with MD mode ($R_{Cav} = 55.2\ nm$), (b) MD source with MD mode, (c) ED source with ED mode ($R_{Cav} = 88.4\ nm$) and (d) MD source with ED mode.

cavity response $a_\alpha(k)$ calculated analytically as shown in Appendix B.

Specifically, Fig. 4(a) and (b) represent the Purcell enhancement caused by the cavity of radius 55.2 nm on an electric and a magnetic dipole source respectively, as a function of the source transition energy. Since at R= 55.2 nm the cavity has a magnetic dipole mode at ~5 eV, we see a significant Purcell enhancement of a magnetic dipole source. Correspondingly, there is a significant de-enhancement of the electric dipolar source, as indicated in panel 4(a). This result points to a potential pathway to selectively enhance dipole-forbidden transitions compared to dipole allowed ones, by tuning a magnetic cavity mode. In Fig. 4(c) and 4(d) we show the Purcell enhancement spectra for the cavity of radius 88.4 nm for which there is an electric dipole resonance at ~5 eV. In this configuration, we observe an enhancement of the electric dipole transition due to the cavity mode, and a corresponding de-enhancement of the magnetic dipole transition. Such configuration can be used to further enhance dipole-allowed transitions and suppress orbital and spin forbidden transitions in quantum emitters.

### B. Distance dependence of matrix elements

The dependence of $M$ on the source-absorber distance originates from the modification of the photon propagator due to the cavity. From equation (15), we can see that at a specific $\alpha$, this distance dependence is:

$$M(\bar{R}) \propto \left[v^{(A)}_{k_S\,\alpha}(\bar{R}) + a_\alpha(k) v^{(A)}_{k_S\alpha[J]}(\bar{R})\right] \quad (20)$$

Here, $v^{(A)}_{k,\alpha_S}(\bar{R}) = \frac{\left\langle ES^{(A)}\,|\tilde{H}_{int}|\; GS^{(A)},1_{k,\alpha_S}(\bar{R})\right\rangle}{\sqrt{\Delta k}}$ where the dependence on $\bar{R}$ is contained in the photon mode. We first discuss a simple physical intuitive picture to explain the various terms that contribute to the distance dependence of $M$ in the presence of a cavity. We consider the case of a transition of a single electron from orbital $\phi_1$ and spin $\chi_1$ to another orbital $\phi_2$ and spin $\chi_2$. The interaction Hamiltonian in the Coulomb gauge and for a single electron can be expressed as $\frac{\hbar e}{2m_0}\left[2\bar{A}_{k,\alpha}(\bar{R}\,)\cdot\bar{\nabla}_1 \; + g\bar{\sigma}\cdot\bar{B}(\bar{R})\right]$. Thus, for a specific $\alpha$, the matrix element can be expressed as

$$M(\bar{R}) \propto 2\langle\phi_2|\left[\bar{A}_{k,\alpha}(\bar{R}\,) + a_\alpha(k)\bar{A}_{k,\alpha[J]}(\bar{R}\,)\right]\cdot\bar{\nabla}|\phi_1\rangle\langle\chi_2|\chi_1\rangle + \langle\phi_2|\left[\bar{B}_{k,\alpha}(\bar{R}\,) + a_\alpha(k)\bar{B}_{k,\alpha[J]}(\bar{R}\,)\right]|\phi_1\rangle \cdot\langle\chi_2|\bar{\sigma}|\chi_1\rangle \quad (21)$$

When the orbitals $\phi_1$ and $\phi_2$ are of opposite parity, e.g. $|\phi_1\rangle = |s\rangle$ and $|\phi_2\rangle = |p\rangle$, and spin is conserved, i.e. $\chi_1 = \chi_2$ ,the first term in equation (21) is the dominant term, resulting in a distance dependence that is approximated by $\langle p|\left[\bar{A}_{k,\alpha}(\bar{R}\,) + a_\alpha(k)\bar{A}_{k,\alpha[J]}(\bar{R}\,)\right]\cdot\bar{\nabla}|s\rangle$. From the expression of $\bar{A}_{k,\alpha}(\bar{R}\,)$, shown in appendix A, we can see that the distance dependence is expected to be a sum of the spherical Hankel function and the spherical Bessel function of type J. If we further assume that the spatial spread of the orbitals is much smaller than the variation of the field, we can take $\left[\bar{A}_{k,\alpha}(\bar{R}\,) + a_\alpha(k)\bar{A}_{k,\alpha[J]}(\bar{R}\,)\right]$ out of the inner product and in this regime we recover the distance dependence of M under the dipole-dipole approximation.

For $|s\rangle$ to $|p\rangle$ spin non-conserving transitions ($\chi_1 \neq \chi_2$) we obtain an expression different from that of the dipole approximation, as the first term in equation (21) vanishes ($\langle\chi_2|\chi_1\rangle = 0$) and only the second term results in a nonzero matrix element ($\langle\chi_2|\sigma_{x/y}|\chi_1\rangle \neq 0$). Further, because the orbital inner product is taken between an s and a p orbital that are orthogonal, the gradient of the B field results in a non-zero matrix element. Thus, the distance dependence of $M$ for spin non-conserving absorption transition is dominated by that of the gradient of the magnetic field. Because the gradient of the Hankel function at near field is much larger than the gradient of the slowly varying Bessel J function, the cavity has a much weaker effect on the distance dependence of M for spin non-conserving transitions, compared to the spin-conserving case. Thus, the range of the spin-conserving transitions is improved more significantly than that of the spin-non conserving transitions in the presence of the photonic cavity.

Equation (20) also provides us with a path to investigate the interference effect between the direct photon propagator ($v^{(A)}_{k_S\,\alpha}(\bar{R})$) and the cavity mediated photon propagator ($a_\alpha(k)v^{(A)}_{k_S\alpha[J]}(\bar{R})$). If the destructive interference condition is met, i.e. $\left[v^{(A)}_{k_S\,\alpha}(\bar{R}) + a_\alpha(k)v^{(A)}_{k_S\alpha[J]}(\bar{R})\right] = 0$ corresponding to the dominant mode $\alpha$, a suppression of the overall energy transfer rate can be achieved. This interference condition provides a way to prevent the energy transfer at specific distances from the source by controlling the energy shift between the cavity mode and the transition energy; it can potentially allow for the suppression of decoherence and trapping processes as well as for the improvement of the stability of certain excited states for memory applications.

For realistic systems a generic transition between many body electron states of multireference character can possess allowed or forbidden orbital parity, and spin conserving and non-conserving components simultaneously. Our framework allows for the description of all possible many-body transitions and all multipolar modes of the photon and hence enables the study of energy transfer between arbitrary systems of defects. For example, we note that for spin non-conserving transitions the second term in equation (21) contributes to the matrix element and the dependence of $M$ on $\bar{R}$ originates from the magnetic field $\left[\bar{B}_{k,\alpha}(\bar{R}) + a_\alpha(k)\bar{B}_{k,\alpha[J]}(\bar{R})\right]$. In many materials containing rare earth ion defects, transitions between orbitals of the same parity are particularly interesting, as they provide magnetic dipole transitions, e.g., Erbium ions implanted in MgO [74,75]. In the following we present illustrative examples of using the energy transfer from a dipolar source to an oxygen vacancy in MgO and other oxides [6].

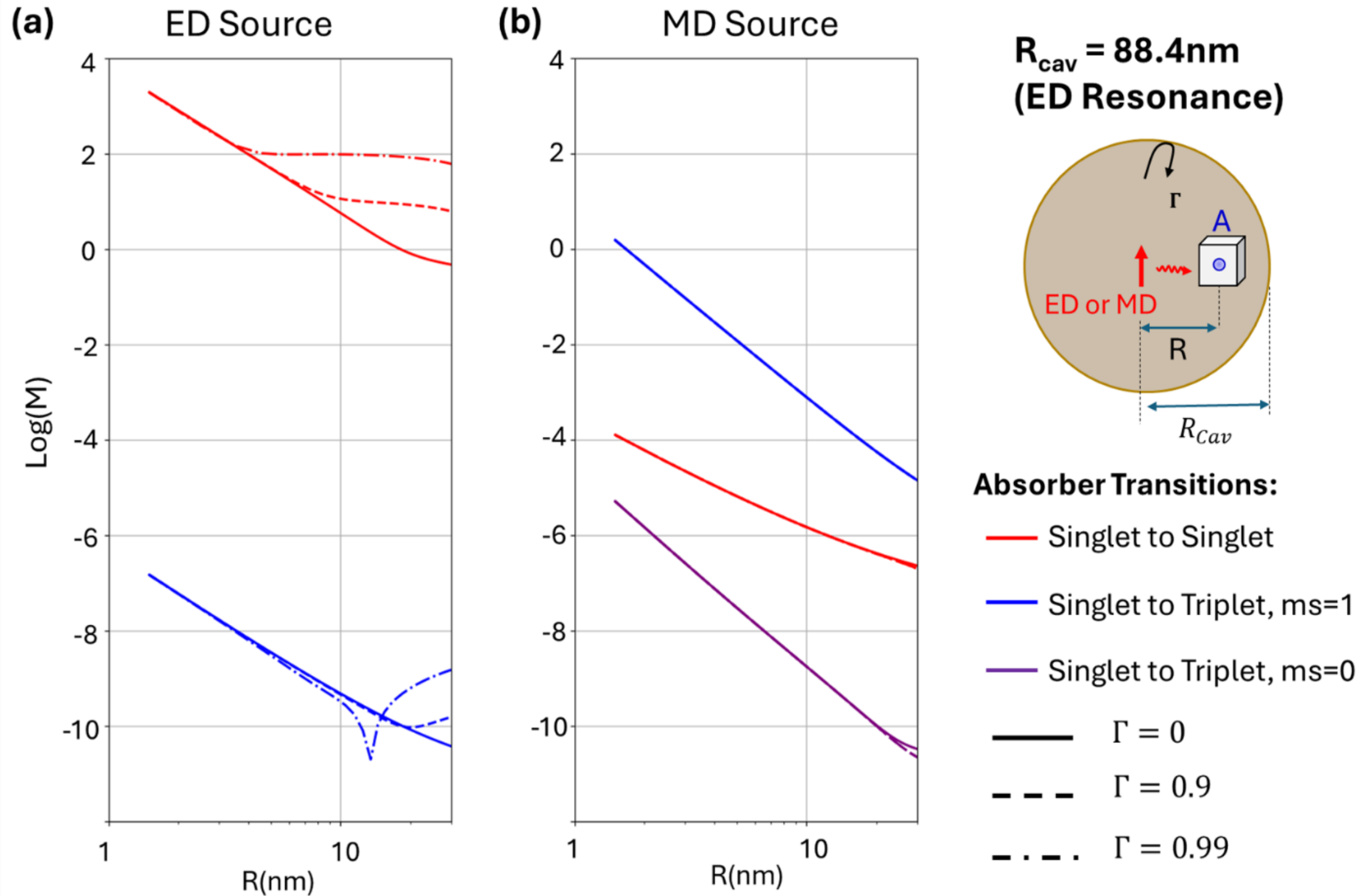

FIG 5. Matrix elements corresponding to the energy transfer process [Eq. (15)] as a function of the distance between an electric/magnetic dipole source and the *F* center in MgO (absorber); the source and the absorber are both embedded in a spherical cavity of radius 88.4 nm (The electric dipole resonance is at 5eV). Panel (a) shows results for an electric dipole source and a cavity wall with reflectivity $\theta = 0$, 0.9, and 0.99. The $\theta = 0$ case mimics an infinite bulk material and serves as a reference. Panel (b) shows results for a magnetic dipolar source and a cavity with wall reflectivity $\theta = 0$, 0.9, and 0.99. The *Y* axis represents *M* in μeV plotted on a log scale.

### C. Energy transfer matrix elements for $V_O$: MgO absorber

Here, we compute the NRET matrix elements using equation (15) for a configuration of the dipolar source at the center of the spherical cavity, and the vacancy in MgO as an absorber, placed at a distance R along the X axis. The results are shown in Fig. 5 and Fig. 6. Specifically, we calculate the energy transfer matrix elements corresponding to the dipole-allowed singlet-to-singlet absorption, and the dipole forbidden singlet-to-triplet absorption. We show below that depending on the choice of the cavity radius and depending on whether the cavity has a magnetic or an electric mode at the emission energy of the source dipole, we obtain drastically different matrix elements.

#### *1. Electric dipole cavity ($R_{Cav}$ = 88.4 nm)*

If the cavity radius is chosen such that the transition energy coincides with the peak of the electric dipole resonance, as per equation (8), the electric dipole type transition matrix element at the source is enhanced while the magnetic dipole transition matrix element is suppressed. In addition, the electric dipole mode excited in the cavity results in a modified photon propagator from the source to the absorber, as shown case. In addition to changes in the dipole allowed transitions, we observe that in the presence of the in equation (15). The dependence $M$ on the source-absorber distance is shown in Fig. 5, where we plot $\log_{10}(|M|)$ as a function of the distance between the source dipole and the F center. The solid lines show cases where Γ (cavity wall reflectivity) is zero, mimicking an infinite bulk material. In this case we obtain the same response as shown in Ref. [6], i.e. the electric dipole source (Fig. 5(a)) results in a dominant spin conserving singlet-to-singlet transition at the F center, while a magnetic dipole source (Fig. 5(b)) leads to a dominant spin non-conserving singlet-to-triplet transition. The dashed lines in Fig. 5(a) show the energy transfer matrix element when the cavity wall reflectivity is chosen as 0.9 and 0.99 (corresponding to Q~20, and Q~400). We find that the enhancement in the energy transfer matrix element from an ED source can reach a significant value of few orders of magnitude and the distance dependence is changed significantly, relative to the homogeneous electric dipole resonance spectrally tuned to the electric dipole source, the energy transfer matrix element of the spin non-conserving transition is significantly modified, both in magnitude and distance dependence. Fig. 5(b) shows the behavior of NRET from a $MD_Z$ source in the cavity with an electric dipole mode. We see that the distance dependence remains mostly unchanged with respect to the case with no cavity (solid line), as expected due to a weak influence of the cavity mode.

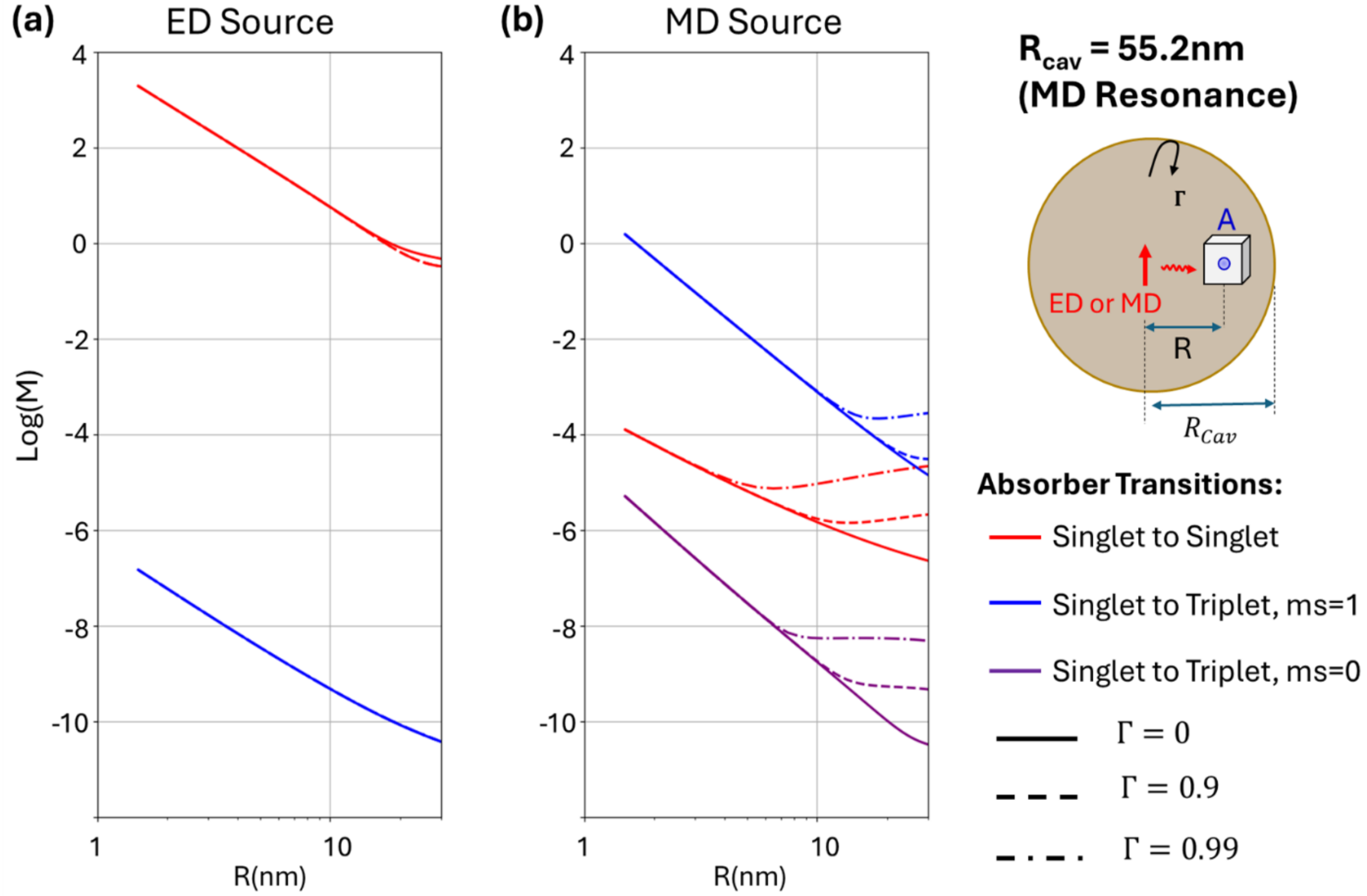

FIG. 6. Matrix elements of the energy transfer process [Eq. (15)] as a function of distance between an electric/magnetic dipole source and the $F$ center in MgO (absorber); the source and the absorber are both embedded in a spherical cavity of radius 55.2 nm (the magnetic dipole resonance is at 5eV). Panel (a) shows results for an electric dipole source and a cavity wall with reflectivity *0* = 0, 0.9, and 0.99. The *0* = 0 case mimics an infinite bulk material and serves as a reference. Panel (b) shows results for a magnetic dipole source and a cavity wall with reflectivity *0* = 0, 0.9, and 0.99. The $Y$ axis represents $M$ in μeV plotted on a log scale.

### *2. Magnetic dipole cavity, $R_{Cav} = 55.2nm$)*

If the cavity radius is chosen to be ~55.2nm so that the magnetic dipole resonance coincides with the transition, and the electric dipole resonance is far detuned, we observe effects that are opposite to those described above. In this configuration, the emission matrix element ($v^{(S)}$) for a magnetic dipole source is enhanced, and electric dipole source is diminished, as described by equation (8). Thus, with increased cavity quality factor, we have an enhancement in energy transfer from a magnetic dipole source, as shown in Fig. 6(b), while the energy transfer from an electric dipole source remains unaffected [Fig. 6(a)]. The enhancement is reflected in both the singlet-to-singlet and singlet-to-triplet absorption, as indicated by the different colors in Fig. 6(b).

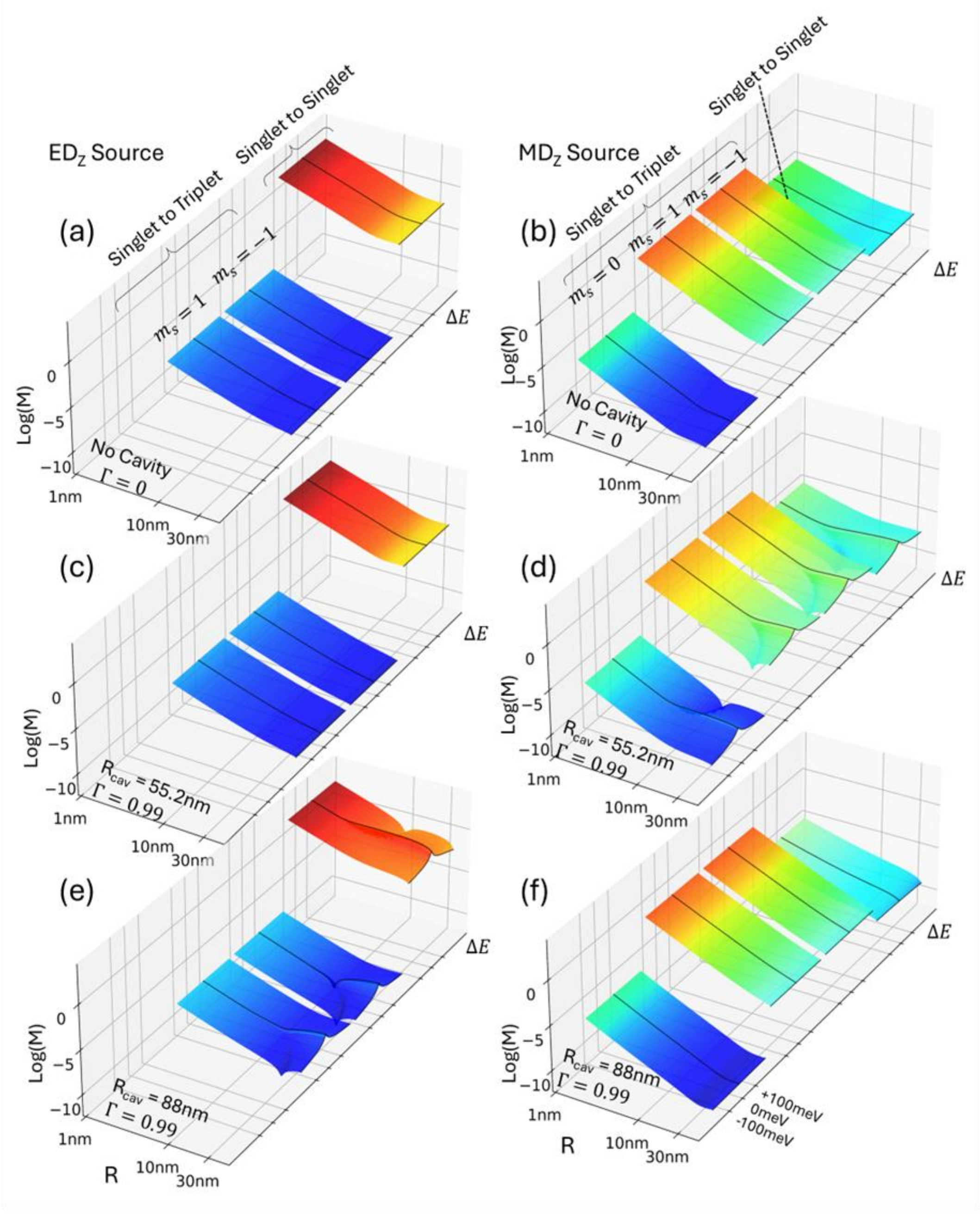

FIG 7. Variation of the matrix element $|M|$ (plotted on a log scale of M in $\mu$eV) as a function of the source-to-absorber distance (R) and energy mismatch ($\Delta E$) between the source transition and the cavity mode, for various configurations. The left column (panels (a), (c), and (e)) show results for an electric dipole (ED) source, and the right column (panels (b), (d), and (f)) for a magnetic dipole (MD) source. The top row (panel (a) and (b)) corresponds to the reference case of no cavity ($\Gamma = 0$). The middle row ((c), (d)) and the bottom row ((e), (f)) show cases corresponding to cavity radii $R_{Cav} = 55.2\ nm$, and $R_{Cav} = 88.4\ nm$ respectively, both for $\Gamma = 0.99$. The transition energy at the source and absorber is fixed at 5eV, and the cavity mode is shifted to result in a spectral mismatch. On each panel, each strip corresponds to a specific transition at the absorber (singlet to singlet, or singlet to triplet with $m_s$ =0, 1, -1). Each strip on the $\Delta E$ axis represent a $\pm 200 meV$ range, and the black line at the center shows the case where the cavity is spectrally matched with the transition energy.

The dependence of $M$ on the source-absorber distance for the spin non-conserving transitions in Fig. 5 and 6 can be now understood with the help of equation (18) to (20). We see from equation (18) that the photon absorption matrix element for the singlet to $m_s = 1$ triplet (blue curve in Fig. 5 and 6) transition is dominated by spin-flip transition between the defect s orbital to the defect p orbital ($\langle p_\downarrow | H_{int} | s_\uparrow, 1_{k,\alpha} \rangle$) which, guided by equation (21), can be also written as $\langle p | [\bar{B}_{k,\alpha}(\bar{R}) + a_\alpha(k)\bar{B}_{k,\alpha[J]}(\bar{R})] | s \rangle \cdot \langle \downarrow | \bar{\sigma} | \uparrow \rangle$. The spin inner product is nonzero for the X and Y components of the Pauli matrices, and thus the gradient of the $B_X$ and $B_Y$ fields provide the dominant distance dependence of M corresponding to the singlet to $m_s = 1$ triplet transition at the F center. On the other hand, for the singlet to $m_s = 0$ triplet (violet curve in Fig. 5 and 6) transition, using equation (21), the absorption matrix element can be expressed as $\langle p | [\bar{B}_{k,\alpha}(\bar{R}) + a_\alpha(k)\bar{B}_{k,\alpha[J]}(\bar{R})] | s \rangle \cdot [\langle \downarrow | \bar{\sigma} | \downarrow \rangle - \langle \uparrow | \bar{\sigma} | \uparrow \rangle]$. This matrix element is only nonvanishing for the Z component of $\bar{\sigma}$. Thus, in this case, the distance dependence of $M$ is provided predominantly by the dependance of the gradient of the $B_z$ field along the Z direction. For both the singlet to $m_s = 0$ triplet and singlet to $m_s = 1$ triplet, since it is the gradient of magnetic field that provides the dominant transition, we see that the photonic cavity does not alter the distance dependence significantly at short distances (R<10nm) compared to the dipole allowed singlet to singlet transition. This is because at near field, the gradient of the Hankel functions dominates over the gradient of the slowly varying Bessel J functions that constitute the cavity mode. Thus, in Fig. 6(b) we see that the radial dependence of the NRET to the singlet-to-singlet transition is rather flat whereas the distance dependence of the NRET to the singlet to triplet transition retains its bulk-like $\sim 1/R^4$ behavior up to ~10nm separation.

The above example demonstrates that all the possible details of the various transitions, spin allowed and forbidden, between many electron states are all combined under a unified framework, which can be used to evaluate the near field resonance energy transfer matrix element $M$ (equation (12), for generic photonic cavities) and equation (15), for specific to spherical cavities). Thus, the framework is immediately applicable to any localized emitters with an arbitrarily complex electronic structure.

### *3. Effect of cavity mode detuning*

In both examples discussed above, we assumed the cavity mode to be spectrally aligned with the transition energy of 5 eV. However, the major advantage of having enhanced NRET using a cavity comes from the fact that the cavity mode can be tuned by various means such as electrooptic or thermal processes. Such processes may provide a microscopic control on individual pairs of emitters for ultra-high density optical memories. Also, controlling energy transfer by tuning the cavity mode provides a way to control coherent transfer between emitters and to entangle operations in quantum memories and networks. To explore these phenomena, we study next the effect of a cavity mode that is detuned with respect to the transition energy.

When the energy mismatch between the cavity mode and the transition is increased, the response $a_\alpha(k)$ shows the resonance profile with symmetric real part and antisymmetric imaginary component as indicated in Appendix B. The overall effect is captured in Fig. 7, where we plot the amplitude of the separation R and the spectral mismatch between the cavity mode and the transition energy (fixed at 5 eV). The NRET corresponding to the singlet-to-singlet absorption and the singlet to triplet absorption with $m_s$=+1, -1, and 0 are plotted in each panel side by side. Figure 7(a) and (b) represent the reference case where $\Gamma = 0$ and the cavity mimics an infinite homogeneous bulk medium. Fig. 7(c) and (d) show a cavity with the $R_{Cav} = 55.2\ nm$ for ED and MD sources, respectively, and Figure 7(e) and (f) show a cavity with $R_{Cav} = 88.4\ nm$ for ED and MD sources.

In Figure 7 panel (d) and (e), the symmetry of the source dipole matches the symmetry of the cavity mode resulting in a peak in the NRET matrix element when the cavity mode is tuned. As the cavity mode is detuned, we see a decrease in the NRET matrix element by ~2 orders of magnitude over ~50 meV detuning for a cavity with $\Gamma = 0.99$. For the cases where the cavity mode symmetry is different than that of the source dipole, i.e. panel (c) and (f), we find a response insensitive to the energy mismatch.

A practically relevant situation occurs when there is destructive interference between the direct coupling term ($v_{k_S\,\alpha}^{(A)}(\bar{R})$) and the cavity-mediated coupling ($a_\alpha(k)v_{k_S\alpha[J]}^{(A)}(\bar{R})$) corresponding to the dominant mode ($\alpha$). When such a condition is achieved, the energy transfer is suppressed, and we find a minimum in the function $|M|$. In Figure 8 we demonstrate this effect. In panel (a) of Fig. 8 we show for clarity the same surface plot also shown in Fig. 7(e) corresponding to the singlet-to-singlet transition. In figure 8 panel (b) we show the section indicated by the dashed black line in (a) corresponding to R=12nm. The dip around $\Delta E \approx 170meV$ indicates an order of magnitude suppression in the value of $|M|$. The reason behind this suppression is illustrated in panel (c) where we plot the real and imaginary components of $M$ at R = 12nm as a function of the energy mismatch $\Delta E$. We see that, because of the asymmetry in the imaginary component of the response $a_\alpha(k)$ [see Appendix B], Im(M) vanishes for a specific $\Delta E$, resulting in a minimum in $|M|$.

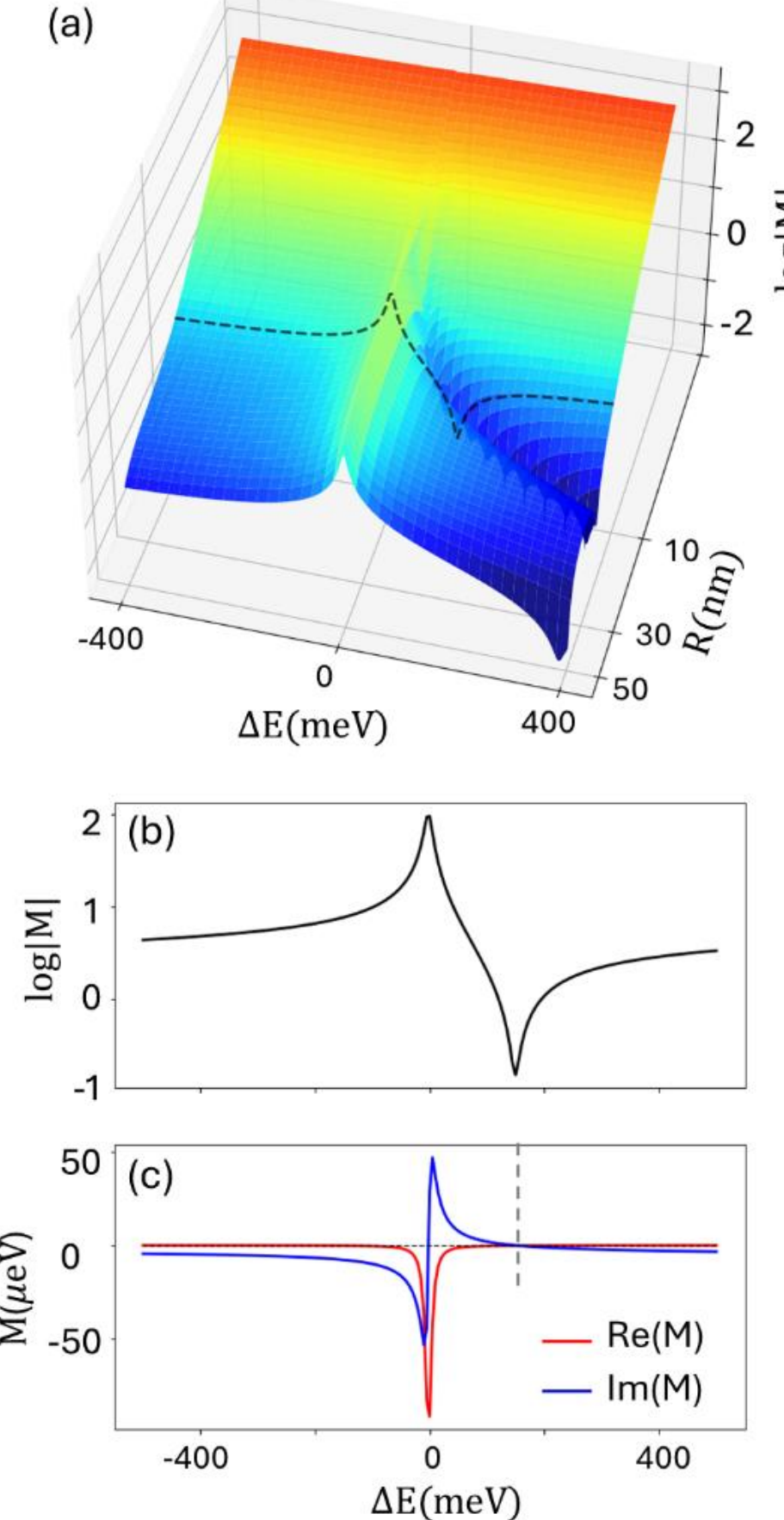

FIG. 8. (a) Surface-plot of log of $|M|$ (see Eq. 20) as a function of the source-absorber distance (R) and energy mismatch between the cavity and the transition (same as shown in Fig. 7(e) and repeated here for clarity). The dashed line corresponds to $R = 12nm$ - also plotted in panel (b). Panel (c) shows the real and imaginary components, respectively. The dip in $|M|$ around $\Delta E \approx 170meV$ corresponds to the destructive interference between the direct coupling term ($v_{k_S\,\alpha}^{(A)}(\bar{R})$) and the cavity-mediated coupling ($a_\alpha(k)v_{k_S\alpha[J]}^{(A)}(\bar{R})$) resulting in the cancellation of the imaginary component of M as indicated by the dashed vertical line in panel (c).

The above results provide further evidence that with a tunable cavity it is possible to control NRET processes between a specific source and a specific absorber. Hence it is a demonstration that the cavity mode can be exploited to enhance the energy transfer rate at ~10nm separation by a factor of 10-100; further, by tuning the cavity mode a suppression of similar order of magnitude can be achieved. Thus, using a suitably designed cavity mode enclosing an ensemble of emitters, desired transitions can be activated, enabling energy transfer in a controllable fashion; they can be deactivated as well- overall providing a long-lasting excited state and preventing decoherence and trapping processes. Such processes are relevant to the design of optical and quantum memory devices.

### D. Incoherent transfer regime

We note that the formulation of NRET developed here (see, e.g. equation (11)), describes the coherent coupling regime where the probability of transfer is represented as $P_{NRET}(t) = \frac{4|M|^2 \sin^2\left(\frac{\Delta\omega t}{2}\right)}{\hbar^2 \Delta\omega^2}$. Thus, with enhancement or suppression of the matrix element M by the cavity, the probability of NRET is $\propto |M|^2$. The effect of the cavity is also present in the incoherent NRET regime, where the emission spectrum of S and the absorption spectrum of A are incoherently broadened, as indicated by the density functions, $\rho_S(\omega)$ and $\rho_A(\omega)$. In that case, in the large t limit, the probability of NRET at time t can be evaluated as a statistical sum given by:

$$P_{NRET}(t) = \frac{2\pi}{\hbar}\int d\omega\, \rho_S(\omega)\rho_A(\omega)\,|M(\omega)|^2 \quad (22)$$

Thus, as long as the cavity mode spectrum overlaps with the joint spectral density of the source and the absorber (i.e. $\rho_S(\omega)\rho_A(\omega)$), the effect of the cavity will be present in the incoherent NRET regime as well. This provides a broad applicability of our framework to systems with a high degree of dephasing such as nanoparticles obtained by solution chemistry and embedded quantum emitters, and room temperature devices.

## IV. CONCLUSIONS

Building on our previous work [6] on first principle NRET in homogeneous media, we presented a generalized framework to predict nonradiative resonant energy transfer processes in a cavity, at arbitrary distances between two realistic solid-state emitters characterized by many-body electronic states. The effect of the cavity on the emitters is described with linear response. The assumption made in our work is that of neglecting the direct exchange of the electrons between the emitters, assuming they are separated far enough to prevent direct orbital overlap. Our approach accounts on the same footing for the two major effects of a general cavity, without any dipole approximation and without any two-level system assumption on the source and the absorber: (1) In the presence of the cavity modes, the transition at the source is dressed, resulting in a negligible Lamb shift and an emission rate enhancement (Purcell effect) in the weak coupling regime. (2) The emitted photons from the source undergo scattering before reaching the absorber, resulting in a modification of the photon propagator from the source to the absorber.

We applied our approach to an exemplary system of absorption into an F center in MgO from a dipole-like source (electric or magnetic) where both the dipole source and the F center absorber are embedded in a spherical cavity mimicking a nanoparticle. By choosing this example we could directly compare with the results of our earlier work [6] in homogeneous media. We investigated two radii of the spherical cavity – 88.4 nm and 55.2 nm, which result in electric dipole and magnetic dipole resonances, respectively. We showed that the electric dipole resonance provides a significant Purcell enhancement to a source electric dipole and significant suppression of a magnetic dipole source, and vice versa. We reported for the first time the comparison between spin conserving and spin non-conserving transitions not addressed in earlier works based on dipole approximation, and we accounted for both the modification of the source oscillator strength and the photon propagation. We showed that the dominant term of the matrix elements associated with orbital forbidden transitions originates from the magnetic field and gradient of the vector potential, whereas that of spin-forbidden transitions originates from the gradient of the magnetic field. More importantly, our framework allows for the first time a general description of source and absorber sites where the transitions are between many-electron states, with an explicit, simultaneous treatment of orbital- and spin-forbidden transitions. Using our framework, we provided new insights into the ways a cavity may be used to control (enhance or suppress) spin non-conserving transitions in a NRET process. Our results indicate that cavity modes can be used to enhance/suppress the NRET processes between emitters by at least a factor of ~100 even when using cavities with moderate Q ~400. Engineering cavity modes by altering the cavity on the micron scale provides a way to control the energy transfer between defects at the ~10 nm scale.

The framework presented here is readily applicable to varied platforms of interest that rely on ensemble of localized quantum emitters embedded in solid state devices over macroscopic separations- including quantum memory, quantum photonics, and optical memory platforms. Because there are no assumptions made on the specific type of the localized emitters, one can investigate optical memories and quantum memories comprising of, e.g., quantum dots, deep level defects, rare earth emitters and native defects. Because we use macroscopic QED to describe the photon propagation, the framework can be applied to coupled emitters over macroscopic and device-level micron-scale distances without any significant increase of computational complexity and without any sacrifice in computational accuracy of light matter interaction at each site.

Interestingly, in the ultra-high density optical memory, a tunable cavity mode can be exploited to enable energy transfer from a specific excited rare earth ion to a specific trap defect. The cavity mode can then be detuned to prohibit further decay or transfer and essentially create long-lived trapped excitations as memory bits. Further, one may control swap operations between two localized emitters [9], often used as communication nodes and memory nodes, which are key to implementing CNOT and other quantum logic operations. In addition, the ability to suppress the NRET process using cavity mode provides a way to suppress decoherence due to spectral diffusion in ensemble emitters in solid state devices, which is essential for good fidelity. Overall, the capability of investigating arbitrary many body transitions paves the way to new avenues to study distribution of entanglement in ordered and random arrays of emitters in a cavity for quantum photonic information processing applications.

## DATA AVAILABILITY

Data that support the findings of this study will be made available through the Qresp [81] curator.

## ACKNOWLEDGMENTS

We thank Prof. Supratik Guha for conceptualization of the ultra-high density optical memory and helpful discussions. We thank Prof. Jorge Sofo and Dr. Yu Jin for helpful discussions. This work was supported by the U.S. Department of Energy (DOE), Office of Science, for support of microelectronics research at the Extreme Lithography & Materials Innovation Center (ELMIC), under contract number DE-AC0206CH11357. We acknowledge the computational resources of the National Energy Research Scientific Computing Center (NERSC), a DOE Office of Science User Facility supported by the Office of Science of the U.S. Department of Energy under Contract No. DE-AC02-05CH11231, and the computational resources of the University of Chicago Research Computing Center (RCC).

## APPENDIX A: Solving response of spherically symmetric cavity

In this work we have assumed a simplified model for a photonic cavity, a sphere surrounded by a coating of reflectivity Γ taken as a free parameter. In this section we outline how such a simplified cavity structure may represent realistic systems based on core-shell structures of nanoparticles. Core-shell nanoparticles can comprise of multiple layers of shells consisting of a metal coating or a high index dielectric material. In general, the response of such structures can be solved using a transfer matrix approach in a spherical basis, as outlined below [76]. In Fig. 9 we show a generic core-shell structure of the spherical nanocavity.

The response can be computed by using Maxwell equations and using the continuity of the tangential E and H fields at the interface. We use the spherical waves as the basis of the photon mode. The transverse component of the vector magnetic potential can be expressed as

$$\bar{A}^{(S)}_{k,\alpha}(\bar{r},t) = \frac{1}{4\pi}\sqrt{\frac{k}{R_{norm}}}\left[\left(\sqrt{\frac{L}{2L+1}}\, g_{L+1}(kr)\,\bar{Y}_{L,L+1,J_z}(\hat{r}) + \sqrt{\frac{L+1}{2L+1}}\, g_{L-1}(kr)\,\bar{Y}_{L,L-1,J_z}(\hat{r})\right)\right] \qquad (A1)$$

In the Coulomb gauge the scalar potential is taken as zero, and the vector magnetic potential is purely a transverse field. The associated electric and magnetic fields are expressed as

$$\bar{E} = i\omega\bar{A}^{(S)}_{k,\alpha} \qquad (A2)$$

and

$$\bar{B} = \bar{\nabla}\times\bar{A} \qquad (A3)$$

We note that $k$ in equation (A1) is given as $k = n_i k_{vac}$, $k_{vac}$ being the vacuum wavevector and $n_i = \sqrt{\epsilon}$ the refractive index, and thus it is valid in any dielectric medium.

Using the above, next the response is readily calculated by expressing the scattered wave as a superposition of radially outward propagating (Hankel type 1) and inward propagating (Hankel type 1) waves:

$$\begin{aligned} &\Gamma e^{2ikR_1}\left|1_{k,\alpha(H2)}\right\rangle, \quad r < R_1 \\ &= b_\alpha^{1,+}\left|1_{k_1,\alpha(H1)}\right\rangle + b_\alpha^{1,-}\left|1_{k_1,\alpha(H2)}\right\rangle, \quad R_1 < r < R_2 \\ &= b_\alpha^{i,+}\left|1_{k_i,\alpha(H1)}\right\rangle + b_\alpha^{i,-}\left|1_{k_i,\alpha(H2)}\right\rangle, R_i < r < R_{i+1} \end{aligned} \qquad (A4)$$

Here $k_i$ is the wavevector of the photon in the i[th] shell. In the outermost shell, the coefficient $b_\alpha$ is set to zero. The boundary conditions are continuity of the tangential E field ($\hat{r}\times\bar{A}$ ) and the tangential magnetic field ($\hat{r}\times\bar{\nabla}\times\bar{A}$ ) at each dielectric interface, i.e. $r = R_1, R_2, \ldots R_n$ resulting in 2n equations. The solution of this matrix equation provides a value for Γ, the effective reflectivity of the nanoparticle shell. For a single dielectric interface, the magnitude of the reflectivity is given by the refractive index contrast. i.e. $|\Gamma| = (n_{in} - n_{out})/(n_{in} + n_{out})$. The phase of Γ can be obtained by including the propagation phase factors from the center to the boundary of the nanoparticles. Thus, for a single dielectric interface, the refractive index contrast limits the value of Γ , and dielectric nanoparticles often result in weak photonic modes with Q<100. Either metal coating or multiple core-shell structure is needed to boost Γ and create a stronger response.

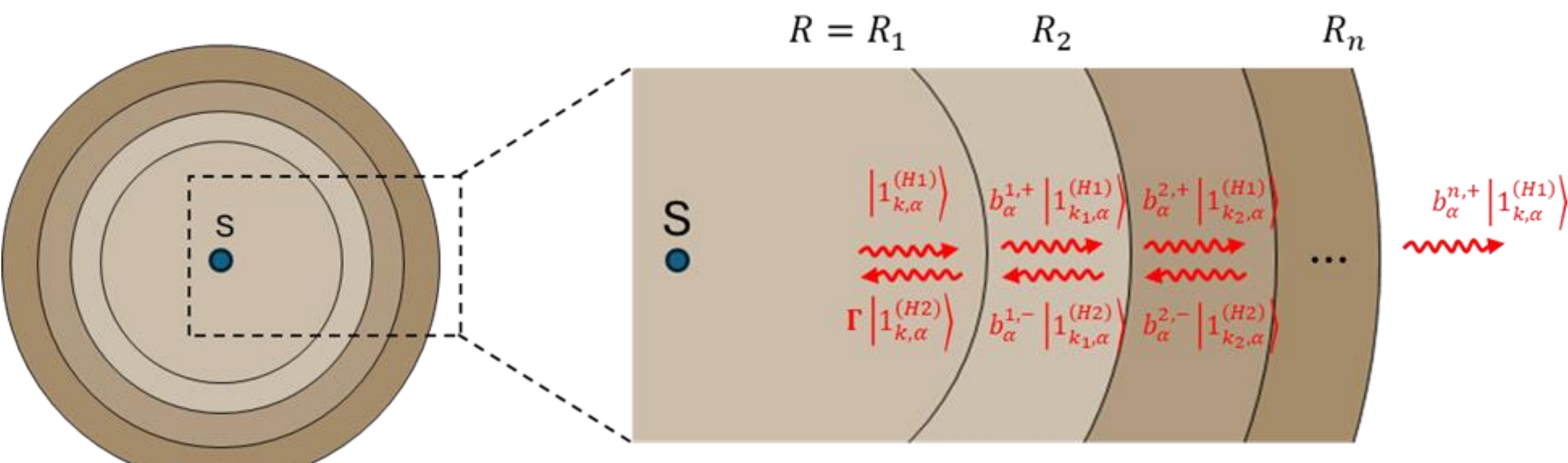

FIG 9. Generic core-shell structure of a spherical cavity and decomposition of scattering of an outward propagating wave ($\left|1_{k,\alpha(H1)}\right\rangle$) in the different dielectric layers. The effective reflectivity into the nanoparticle core is given by Γ.

Once Γ is obtained using the above, the reflected wave $\Gamma \left|1_{k,\alpha(H2)}\right\rangle$ produces a standing wave, as a result of multiple scattering. This provides us with a way to calculate the cavity response $a_\alpha(k)$ using the equation

$$a_\alpha(k) = \Gamma + \Gamma^2 + \Gamma^3 \ldots = \frac{\Gamma}{1-\Gamma} \qquad (A5)$$

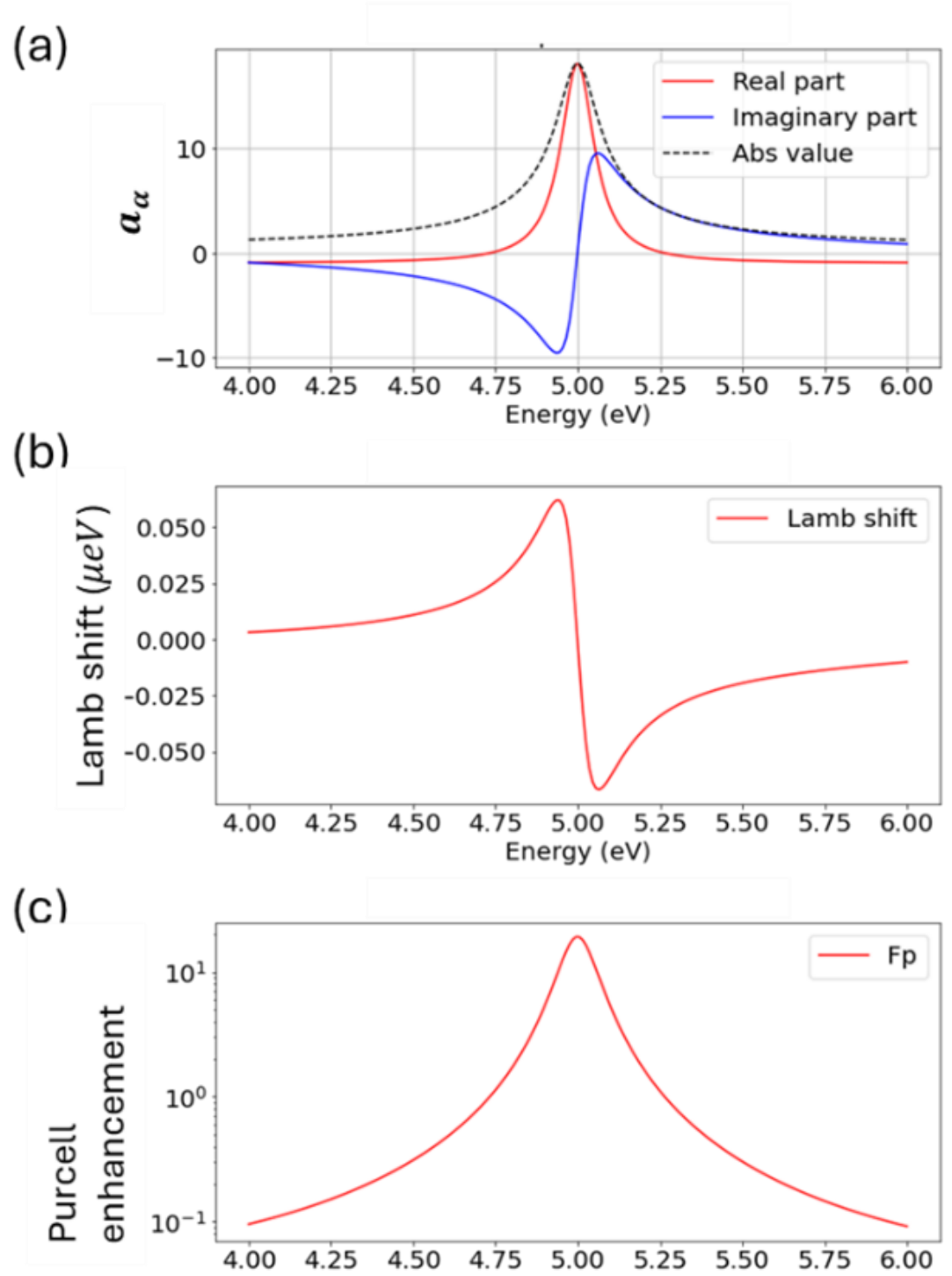

FIG 10. (a) Spectrum of the response of an ED cavity (radius R = 88.4 nm) and wall reflectivity Γ = 0.9. Panel (b) and (c) show the Lamb shift and the Purcell enhancement spectra, respectively, for a 1 oscillator strength dipole source.

## APPENDIX B: Lamb shift and Purcell enhancement in the weak coupling regime

In Fig. 10(a) we show the complex spectrum of the cavity response function for a spherical cavity of radius 88.4 nm and $\alpha$ = {L = 1, $J_z$ = 0, P = -1} corresponding to a magnetic dipole mode, and Γ=0.9. As expected, the real part exhibits symmetric behavior, and the imaginary part shows an antisymmetric behavior around the resonance. From equation (6), the self-energy can be expressed as

$$\Sigma \approx \frac{i\pi n_i}{\hbar c}\left|v^{(S)}_{k_S\alpha}\right|^2 \left(1 + a_\alpha(k_S)\ \right) \text{ at } \alpha = ED \text{ mode} \qquad (B1)$$

Thus, the Lamb shift ($Re(\Sigma)$) is proportional to the imaginary part of $a$. This is shown in Fig. 10(b) which shows the Lamb shift for a dipole source of unit oscillator strength. Note that the amount of the shift is negligible and, in most cases, can be ignored. The real part of $a_\alpha(k)$ contributes to the imaginary component of the self-energy and results in an enhanced decay, reflected in the Purcell enhancement spectrum shown in Fig. 10(c).

## APPENDIX C. Absorber transitions at $V_O$: MgO

In this work we used Kohn-Sham density functional theory and the Quantum Espresso code to obtain the single electron orbitals involved in the singlet to singlet and the singlet to triplet transitions in the F center. We used the SG-15 norm conserving Vanderbilt pseudopotential [77] and both PBE [78] and dielectric dependent hybrid (DDH)[79] functional resulting in very close orbitals.

Figure 11(a) and (b) show the energy levels and the orbitals corresponding to the mid-gap s-orbital and the above conduction band minimum p type defect orbital that dominantly participate in the optical absorption of the F center. The eigenstates can be further refined using many-body perturbation theories and quantum embedding theories, since the framework shown here extends to general many-electron states with ease. However, for simplicity such further refinement in the electronic structure part is not shown in this work.

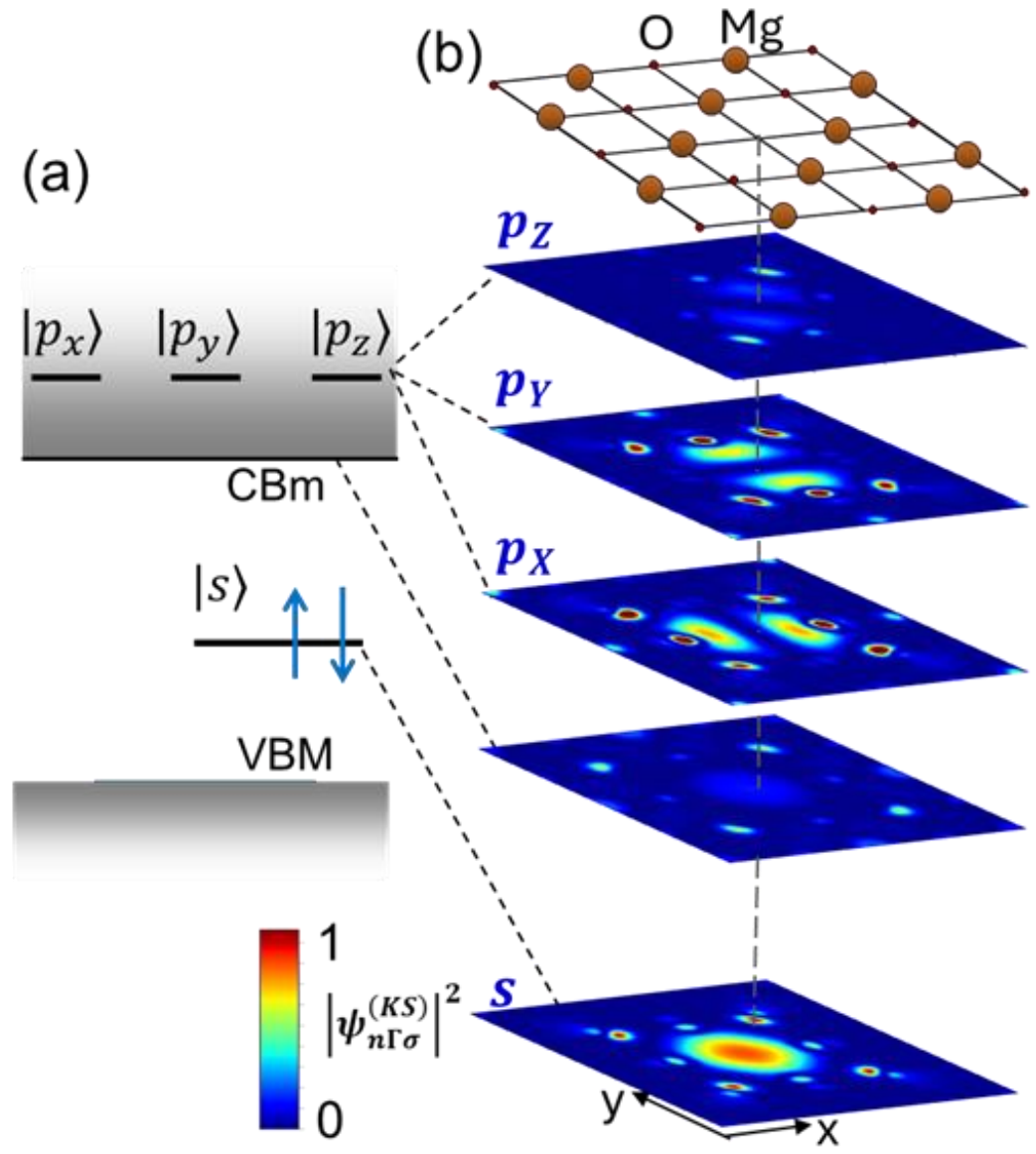

FIG 11. (a) Kohn-Sham energy (schematic) of the mid-gap s orbital and the above-CBm (conduction band minimum) p-orbital that participate in the optical absorption at the F center, and (b) the section of the Kohn-Sham orbitals from DFT calculations.

**APPENDIX D. Angle-dependent behavior**

In Figure 7 we have shown the dependence of M (Eq. 20) on the source-absorber distance corresponding to the specific situation where the displacement between the dipolar source and the absorber is along a fixed direction perpendicular to the dipole moment of the source. In this section, we report on the dependence of M on the source-dipole orientation. The impact of the molecular orientation of the source and absorber on the energy transfer rate has been of interest in plasmonic systems [80] and it has been shown that sharp enhancement of energy transfer can be enabled at specific angles. In our case, since the source is located at the center of the spherical cavity, the effect of the orientation of the dipole source on the transfer rate is not significant. This is highlighted in figure 12 and 13- corresponding to the cavity with an electric dipole resonance ($R_{Cav} = 88.4nm$) and magnetic dipole resonance ($R_{Cav} = 55.2$); in both cases the cavity mode is tuned to $\omega_S$(i.e. $\Delta E = 0$). Panel (a), (b) and (c) in both figures show the case corresponding to an electric dipole source as a function of the distance (R) and angle ($\theta$) for $\Gamma = 0$, 0.9, and 0.99 respectively. The panels (d),(e) and (f) show the same quantities for a magnetic dipole source. We find a behavior that is consistent for most angles, and that the direction at which the absorber is from the source does not play a critical role, consistent with the spherically symmetric configuration represented in this work, and can potentially be important for more complex emitter-cavity configurations and non-spherical cavity geometries.

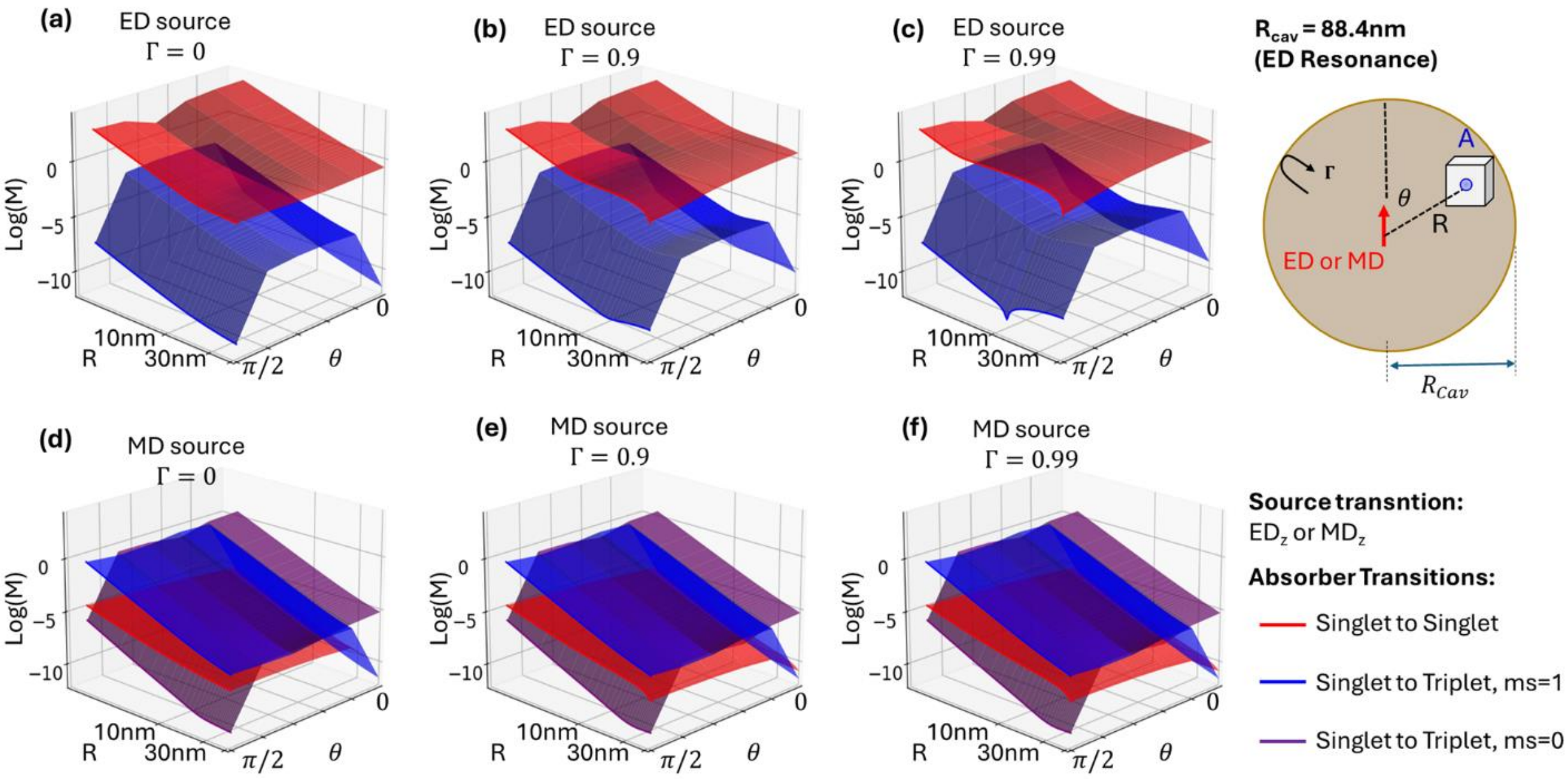

FIG 12. Variation of the matrix element $|M|$ (see Eq. 20, plotted on a log scale of M in $\mu$eV) as a function of the source-to-absorber distance (R) and direction ($\theta$) (shown in inset) corresponding to a cavity with electric dipole resonance (R = 88.4nm) and cavity resonance tuned to the transition energy ($\Delta E = 0$). Panel (a), (b) and (c) show the case for an ED source and $\Gamma = 0,0.9,0.99$ respectively. The red, blue, and violet surfaces correspond to the singlet to singlet, singlet to $m_s$ =0 and $m_s = 1$ triplets. Panels (d),(e), and (f) show the equivalent for a magnetic dipole source.

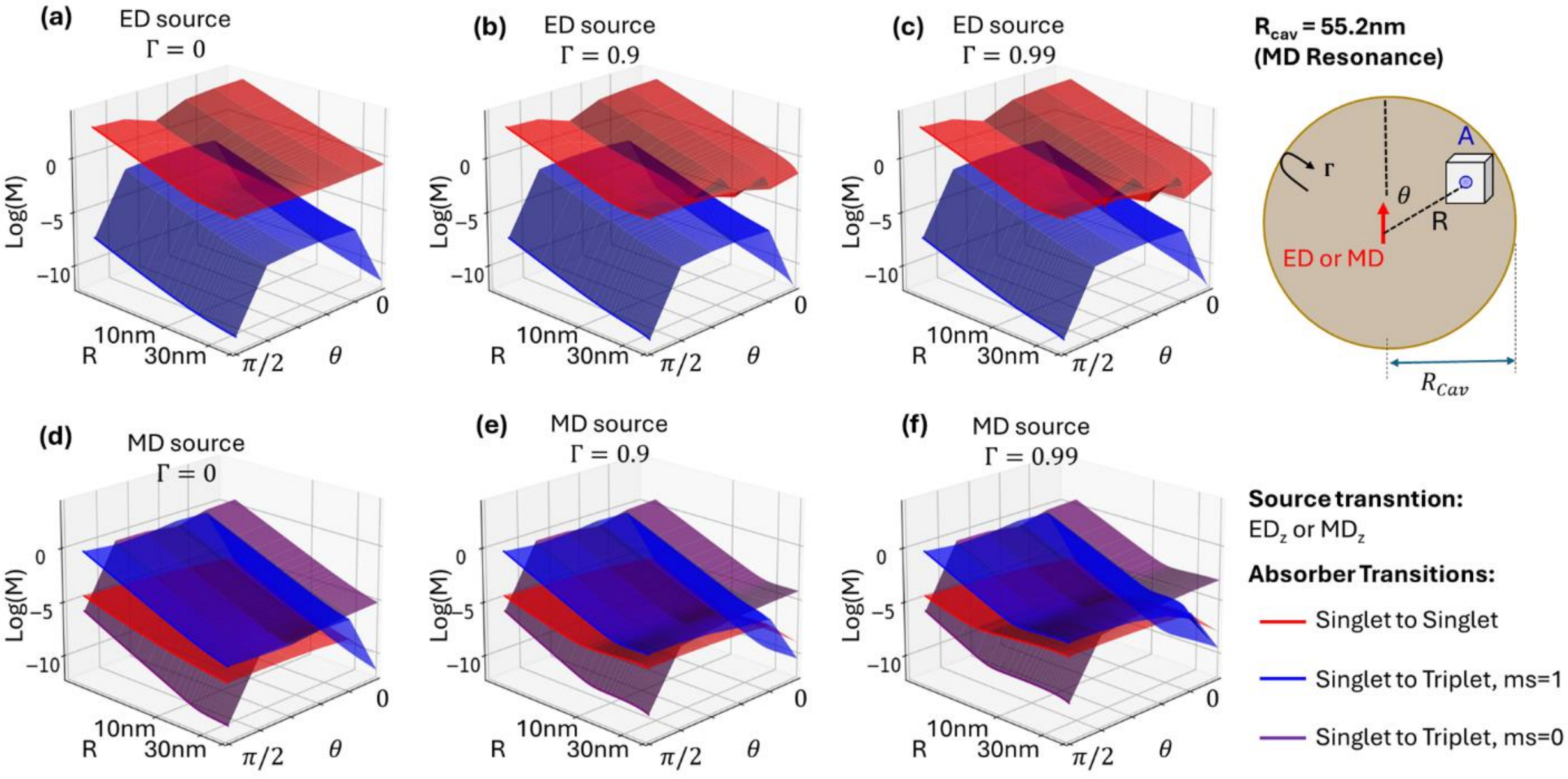

FIG 13. Variation of the matrix element $|M|$ (see Eq. 20; plotted on a log scale of M in $\mu$eV) as a function of the source-to-absorber distance (R) and direction ($\theta$) (shown in inset) corresponding to a cavity with magnetic dipole resonance (R = 55.2nm) and cavity resonance tuned to the transition energy ($\Delta E = 0$). Panel (a), (b) and (c) show the case for an ED source and $\Gamma = 0,0.9,0.99$ respectively. The red, blue, and violet surfaces correspond to the singlet to singlet, singlet to $m_s$ =0 and $m_s = 1$ triplets. Panels (d),(e), and (f) show the equivalent for a magnetic dipole source.